\documentclass[12pt]{article}

\begin{document}
\parskip 1ex
\setcounter{page}{1}
\oddsidemargin 0pt
\evensidemargin 0pt
\topmargin -40pt
%
\newcommand{\be}{\begin{equation}}
\newcommand{\ee}{\end{equation}}
\newcommand{\ba}{\begin{eqnarray}}
\newcommand{\ea}{\end{eqnarray}}
\newcommand{\bea}{\begin{eqnarray}}
\newcommand{\eea}{\end{eqnarray}}
\def\a{\alpha}
\def\b{\beta}
\def\g{\gamma}
\def\G{\Gamma}
\def\d{\delta}
\def\e{\epsilon}
\def\z{\zeta}
\def\h{\eta}
\def\th{\theta}
\def\k{\kappa}
\def\l{\lambda}
\def\L{\Lambda}
\def\m{\mu}
\def\n{\nu}
\def\x{\xi}
\def\X{\Xi}
\def\p{\pi}
\def\P{\Pi}
\def\r{\rho}
\def\s{\sigma}
\def\S{\Sigma}
\def\t{\tau}
\def\f{\phi}
\def\F{\Phi}
\def\c{\chi}
\def\w{\omega}
\def\W{\Omega}
\def\de{\partial}

\def\pct#1{(see Fig. #1.)}

\begin{titlepage}
\hbox{\hskip 12cm ROM2F-01/24  \hfil}
\hbox{\hskip 12cm hep-th/0107090 \hfil}
\vskip 1.4cm
\begin{center}  {\Large  \bf   Geometric \ Couplings \ and   \vskip
.6cm  Brane \ Supersymmetry \ Breaking}

\vspace{1.8cm}
 
{\large \large  Gianfranco Pradisi \ and \ Fabio Riccioni}
\vspace{0.6cm}

{\sl Dipartimento di Fisica, \ \ Universit{\`a} di Roma \ ``Tor Vergata'' \\
I.N.F.N.\ - \ Sezione di Roma \ ``Tor Vergata'', \\ Via della Ricerca
Scientifica , 1 \ \ \ 00133 \ Roma \ \ ITALY}
\end{center}
\vskip 1.0cm

\abstract{Orientifold vacua allow the simultaneous presence of 
supersymmetric bulks, with one or more gravitinos, and
non-supersymmetric combinations of BPS branes. This ``brane
supersymmetry breaking''
raises the issue of consistency for the resulting
gravitino couplings, and Dudas and Mourad recently provided convincing 
arguments to this effect for the ten-dimensional $USp(32)$ model. 
These rely on a non-linear realization of local supersymmetry 
{\it \`a la} Volkov-Akulov, although no gravitino mass term is 
present, and the couplings have a nice geometrical 
interpretation in terms of ``dressed'' bulk fields, aside from 
a Wess-Zumino-like term, resulting from the supersymmetrization
of the Chern-Simons couplings. Here we show that {\it all} couplings 
can be given a geometrical interpretation, albeit in the
dual 6-form model, whose bulk includes a Wess-Zumino term, 
so that the non-geometric ones are in fact demanded by the geometrization 
of their duals. 
We also determine the low-energy couplings for 
six-dimensional (1,0) models with brane supersymmetry breaking. Since
these include both Chern-Simons and Wess-Zumino terms, 
only the resulting field equations are geometrical, aside from
contributions due to vectors of supersymmetric sectors.} 

\vskip 1.0cm
\begin{center}
{( July, \ 2001 )}
\end{center}
\vfill
\end{titlepage}
\makeatletter
\@addtoreset{equation}{section}
\makeatother
\renewcommand{\theequation}{\thesection.\arabic{equation}}
\addtolength{\baselineskip}{0.3\baselineskip} 

\section{Introduction}
Orientifold vacua \cite{cargese,dudrev} allow the simultaneous presence of 
supersymmetric bulks, with one or more gravitinos, and
non-supersymmetric combinations of BPS branes. The resulting 
``brane supersymmetry breaking'' can be realized in stable
configurations, in ten dimensions with 
only anti-$D9$ (${\bar{D}}9$) branes \cite{sugimoto}, 
and in six and four dimensions, up to 
T-dualities, 
with tachyon-free combinations of $D9$ branes and anti-$D5$ (${\bar{D}}5$) 
branes \cite{bsb}. Since this is only one of the options offered
by this class of models for the breaking of supersymmetry, a 
fundamental issue in attempting to relate string
theory to low-energy physics, it is instructive to briefly review our 
current knowledge in this respect.
 
In perturbative string vacua, one has actually 
four options for supersymmetry breaking. The first is to break
supersymmetry from the start, 
so that no gravitinos are present, and the resulting
models, descendants of the type-0 models of \cite{dhsw}, have in general
tachyonic modes \cite{bianchi90}, although a special Klein-bottle projection, 
suggested by
the WZW constructions of \cite{pss}, leads to the $0'$B
model \cite{as95}, that is free of tachyons, a property shared by its
compactifications \cite{carlo} and neatly rooted in 
its brane content \cite{dm1}.
The second option is the Scherk-Schwarz 
mechanism \cite{ss}, in which the breaking, induced by
deformed harmonic expansions in the internal space, is at the 
compactification scale.  In this setting, widely studied in the context
of models of oriented closed strings \cite{ssclosed}, the
 presence of branes allows the new 
option of correlating the Scherk-Schwarz deformations
to the brane geometry, giving rise, in particular, to the phenomenon of
``brane supersymmetry'', whereby one or more residual 
global supersymmetries are left, to lowest order, for the brane modes
\cite{ssopen}.
The third option, magnetic deformations \cite{ft}, resorts to the
different magnetic moments of the various fields to induce supersymmetry
breaking  \cite{bachas}, again at the compactification scale, 
but the resulting vacua, that have also T-dual descriptions in terms
of branes at angles \cite{douglas}, generally contain tachyons \cite{berlin}, 
aside from some special instanton-like stable configurations that
recover supersymmetry, albeit with gauge groups of reduced rank \cite{aads}.
Finally, one has the option of brane supersymmetry breaking 
\cite{sugimoto,bsb}, made possible by the presence
of two types of $O$-planes. Together with the conventional $O_+$,
with negative tension and negative R-R charge, there are indeed additional
BPS objects, the  $O_-$ planes,
with positive tension and positive R-R charge, and while the two can
coexist in supersymmetric Klein-bottle projections, the saturation
of the $O_-$ charge requires the presence of anti-branes, with the
result that supersymmetry is broken on the latter {\it at the string
scale}. It is the rigidity of the breaking scale, together with some
special features of the resulting low-energy effective field theories, 
that typically do not allow a gravitino mass term,
that makes the explicit construction of the goldstino couplings
quite interesting in this case. 

Dudas and Mourad \cite{dm2} have recently shown that, in 
the simplest model with brane supersymmetry breaking, the 
$USp(32)$ model of \cite{sugimoto}, the low-energy gravitino
couplings reflect a non-linear realization of local supersymmetry 
{\it \`a la} Volkov-Akulov, 
along the lines of \cite{samwess}, and their work
is the starting point for our considerations. Let us stress that, while
all branes, including the supersymmetric ones, result in the non-linear
realization {\it \`a la} Born-Infeld of the supersymmetries broken by their 
presence, here one arrives at a complete breaking, and the peculiarity
with respect to lower-dimensional settings for the super-Higgs mechanism
is the absence of a gravitino mass term. This feature is common to the
case analyzed in \cite{dm2} and to the lower-dimensional models of \cite{bsb},
that we shall also discuss in this paper. Actually, all these
configurations, even the supersymmetric ones, can accommodate 
additional brane-antibrane pairs of identical
dimensions, that are to be
spatially separated in order to lift the resulting tachyons. These 
additional pairs provide in their own right additional ways 
to realize brane supersymmetry breaking, but have clearly potential 
tachyon instabilities \cite{abg} for their geometric moduli, in view of 
the mutual attraction of identical branes
and antibranes. We shall thus confine our attention to 
the ``minimal'' configurations of \cite{sugimoto,bsb} demanded by tadpole 
cancellation,
although the other pairs could be described along similar lines. 
Still, we should mention that non-minimal brane-antibrane configurations 
are also quite interesting, and are currently the object of a considerable 
activity as a string setting for brane-world extensions of the Standard Model \cite{berlin,ibanez}.

All models with brane supersymmetry breaking contain a candidate goldstino
among their brane modes, and in \cite{dm2} Dudas and Mourad indeed
constructed the low-energy couplings of the goldstino for 
the ten-dimensional $USp(32)$ Sugimoto model of \cite{sugimoto} up to quartic 
fermionic terms. These were all shown to be of a geometric nature,
being induced by the dressing
of bulk fields with additional terms depending on the goldstino in all
their couplings to the non-supersymmetric brane matter,
aside from some Wess-Zumino-like terms resulting from the supersymmetrization 
of the Chern-Simons couplings. The geometric nature of the
dressing implies that
non-linear supersymmetries of the matter sector take the form of
gaugino-dependent general coordinate transformations.
In this paper, we extend the work of \cite{dm2}, showing
that, up to quartic fermionic terms, the {\it{whole}} low energy effective
Lagrangian of the Sugimoto model, including the Chern-Simons couplings, 
has a geometric nature when expressed 
in terms of the dual $6$-form gauge field, rather than of the more familiar 
$2$-form.  The starting point in this case is thus the
low-energy supergravity built long ago by Chamseddine \cite{chamseddine},
rather than the model of \cite{bddv}. The ten-dimensional Chern-Simons 
terms become in this way higher derivative couplings that, as such, 
do not appear in the low-energy effective action, while a
Wess-Zumino term must be added, and this can be simply ``geometrized'' 
dressing the six-form along the lines of \cite{dm2}.  We also 
extend the analysis of the low-energy effective action to 
six dimensional models with brane supersymmetry breaking.
The starting point in this case is provided by the low-energy (1,0)
effective actions of \cite{ggs,ggs2,ggs3,rs2,fabio}. These, however, include both
Wess-Zumino and Chern-Simons couplings for the gauge fields, and as
a result have the subtle feature of embodying reducible gauge anomalies to 
be canceled by fermion loops. This peculiar feature,
not present in the earlier constructions of \cite{nsez1} motivated by
perturbative heterotic strings, links these constructions to the
Wess-Zumino conditions for the anomalies, with the end result that
many familiar properties of current algebra find in this case an
explicit local realization. In order to 
write a covariant action for the resulting (anti)self-dual 3-forms,
we shall resort to the method of Pasti-Sorokin-Tonin \cite{pst}, both 
for the (1,0) couplings of vector and tensor multiplets of \cite{ggs}
and for the additional hypermultiplet couplings obtained by one of
us in \cite{fabio} completing the results of \cite{nsez2}. The remaining 
couplings are determined requiring that supersymmetry be
non-linearly realized as in the ten-dimensional case, but the
simultaneous presence of Chern-Simons and Wess-Zumino terms produces
a novel effect.  Indeed, while the action is still determined by the 
underlying geometrical structure, as is often the case with Wess-Zumino 
terms, only the field equations are geometrical in this case, aside from 
anomalous terms that arise in the presence of vectors from both 
supersymmetric and non-supersymmetric sectors.

The paper is organized as follows. In Section 2 we review some 
basic facts on type-I vacua with
brane supersymmetry breaking, referring to the Sugimoto 
model \cite{sugimoto} and to the simplest lower-dimensional model with
$D9$ and ${\bar{D}}5$ branes, the $T^4 / Z_2$ orientifold
of \cite{bsb}.  In Section 3 we review the 
low-energy effective couplings built by Dudas and Mourad \cite{dm2}
for the ten-dimensional model and exhibit their geometric nature 
in terms of the 6-form potential. Section  4 is devoted to the
six-dimensional non-linear realizations, and in order to make the paper 
self-contained, it includes a brief review of
the results of \cite{ggs,ggs2,ggs3,rs2,fabio}, limitedly to the terms at most
quadratic in the fermions needed for our construction. 
Finally, Section 5 contains our conclusions.

\section{Models with brane supersymmetry breaking}

In the following sections we shall describe the low-energy goldstino
couplings for what are essentially the two classes of string realizations
of brane supersymmetry breaking. The first, defined in ten-dimensional
space time, is the Sugimoto model of \cite{sugimoto}, while the second
is a peculiar $T^4/Z_2$ orientifold, where the Klein-bottle projection is
altered in a way reminiscent of the WZW constructions in \cite{pss}.
The ten-dimensional model involves ${\bar{D}}9$ branes and $O_-$
planes, and results from an unoriented projection obtained combining
the world-sheet parity $\Omega$ with $(-1)^{F_{s}}$, where $F_s$ denotes
the space-time fermion number. Still, the
resulting Klein-bottle projection is identical to the usual one for the
Green-Schwarz $SO(32)$ superstring \cite{gs}
\be
{\cal K}= \ \frac{1}{2} \ (V_8 - S_8)   \ ,
\ee
here expressed, as in \cite{bianchi90}, in terms of level-one $so(8)$
characters while leaving the modular integration implicit.
The difference between the two models lies in the open sector, 
and in particular in the M\"obius amplitude, that reflects the
relative values of tension and charge for branes and $O$-planes.
Thus, for the $SO(32)$ model
\bea
{\cal A}= \ \frac{N^2}{2} \ (V_8 - S_8)  \  , \nonumber \\
{\cal M}= \ - \frac{N}{2} \ (\hat{V}_8 - \hat{S}_8) \ ,   
\label{gs}
\eea
while for the $USp(32)$ model
\bea
{\cal A}= \ \frac{N^2}{2} \ (V_8 - S_8) \  , \nonumber \\
{\cal M}= \  \frac{N}{2} \ (\hat{V}_8 + \hat{S}_8) \ ,  
\label{sugi}
\eea
again in the compact notation of \cite{bianchi90},
with real ``hatted'' characters for the M\"obius amplitudes.
In both cases R-R tadpole cancellation fixes $N=32$, but while in
the former 
32 $D9$ branes cancel the $O_+$ background charge and tension, in
the latter 32 ${\bar{D}}9$ branes cancel only the opposite $O_-$
charge, leaving a resulting dilaton tadpole. The massless
spectrum contains a vector in the adjoint representation
of $USp(32)$ and a spinor in the antisymmetric representation.  This,
however, is not irreducible, but can be decomposed into the $\bf{495}$ 
and a singlet: it is exactly this
singlet that plays the role of the goldstino in the low energy 
effective action.  
Supersymmetry is broken in the open sector, but the spectrum is
free of anomalies, and the dilaton tadpole that survives signals the
impossibility of vacuum configurations with the maximal ten-dimensional
symmetry. Vacuum configurations with a lower, nine-dimensional, symmetry 
are however possible, as shown in \cite{dm3}.

Lower dimensional models with brane supersymmetry breaking were built in
\cite{bsb}.  As anticipated, they result from Klein-bottle projections 
modified in a way reminiscent of what done in WZW models in \cite{pss}.
In $Z_2$ orbifold compactifications, and in general for 
orbifolds with order-two group generators, one has the option 
of antisymmetrizing some twisted sectors or, equivalently, of inverting
tensions and charges for the corresponding $O5$-planes.  This requires
that $D9$ branes be accompanied by suitable stacks of ${\bar{D}}5$ branes,
with a resulting brane supersymmetry breaking.
In the six-dimensional $T^4/Z_2$ orientifold, the
torus partition function 
\ba  
{\cal T} &=& \frac{1}{2} \ | \ V_4 O_4 + O_4 V_4 - S_4 S_4 - C_4 C_4 \ |^2 \ \Lambda + 
\frac{1}{2} \ | \ V_4 O_4 - O_4 V_4 + S_4 S_4 - C_4 C_4 \ |^2 \ 
{\biggl| \ \frac{2 \eta}{\theta_2} \ \biggr|}^4 
\nonumber \\
&+& \frac{1}{2} \ | \ O_4 C_4 + V_4 S_4 - S_4 O_4 - C_4 V_4 \ |^2 \ {\biggl|\frac{2 \eta}
{\theta_4}\ \biggr|}^4  \nonumber \\
&+&
\frac{1}{2} \ | \ O_4 C_4 - V_4 S_4 - S_4 O_4 + C_4 V_4 \ |^2 \ {\biggl| \ \frac{2 \eta}{\theta_3} \ \biggr|}^4 
\label{toro}
\ea  
can be expressed in terms of the Narain lattice $\Lambda$ and of the
$so(4)$ level-one characters $O_4$, $V_4$, $S_4$ and $C_4$, again omitting the modular 
measure and the inert contributions. Two choices for
the unoriented projection compatible with the crosscap 
constraint of \cite{pss} are described by
\ba 
{\cal K} &=& \frac{1}{4} \biggl\{ \ ( \ V_4 O_4 - O_4 V_4 + S_4 S_4 - C_4 C_4 \ ) \ ( P + W ) \nonumber \\
&+& 2 \epsilon \times 16 \ ( \ O_4 C_4 + V_4 S_4 - S_4 O_4 - C_4 V_4 \ ) \ 
{\biggl( \ \frac{\eta}{\theta_4}\biggr)}^2 \ \biggr\} \ , 
\label{klein}
\ea
where $P$ ($W$) indicates the momentum (winding) lattice sum and
$\epsilon = \pm 1$.  At the massless level, $\epsilon = 1$
gives $N=(1,0)$ 
supergravity with $1$ tensor multiplet and $20$ hypermultiplets, 
while $\epsilon = - 1$ gives
$N=(1,0)$ supergravity with $17$ tensor multiplets and $4$
hypermultiplets. These closed spectra are both supersymmetric, but
the latter projection introduces $O9_+$ planes and $O5_-$ planes, and
this leads to an open sector with brane supersymmetry breaking.
This is clearly spelled by the massless contributions to the transverse
Klein-bottle amplitude, that can be read from
\be
\tilde{\cal K}_0 = \frac{2^5}{4} \biggl\{ \ ( \ V_4 O_4 - C_4 C_4 \ ) \ \biggl( \ \sqrt{v}  +
\epsilon\, 
\frac{1}{\sqrt{v}} \ \biggr)^2 + ( \ O_4 V_4 - S_4 S_4 \ ) \ \biggl( \ \sqrt{v}  -
\epsilon \,
\frac{1}{\sqrt{v}} \ \biggr)^2 \ \biggr\} \ , \label{a6}
\ee 
where the reflection coefficients
are interchanged in the two cases.  For $\epsilon=-1$, all
R-R tadpoles can be canceled by 32 $D9$ branes and 32
${\bar{D}}5$ branes, since the latter indeed 
revert all the 9-5 R-R contributions to the transverse-channel 
annulus amplitude.  The result is 
\ba
\tilde{\cal A} &=&  \frac{2^{-5}}{4}  \ \biggl\{ \ 
( \ V_4 O_4 + O_4 V_4 - S_4 S_4 - C_4 C_4 \ ) \biggl( \ N^2 v
W  +
\frac{D^2 P}{v} \ \biggr) \nonumber \\
&+& 2 N D \ ( \ V_4 O_4 - O_4 V_4 - S_4 S_4 + C_4 C_4 \ ) \ {\biggl(\frac{2
\eta}{\theta_2}\ \biggr)}^2  \nonumber \\ 
&+&  16 \ ( \ O_4 C_4 + V_4 S_4 - S_4 O_4 - C_4 V_4 \ ) \biggl(
R_N^2 + R_D^2 \ \biggr){\biggl(\frac{
\eta}{\theta_4}\biggr)}^2 \nonumber \\
&+& 8 R_N R_D \ ( \ V_4 S_4 - O_4 C_4 - S_4 O_4 + C_4 V_4 \ ) \ {\biggl(\frac{
\eta}{\theta_3}\ \biggr)}^2 \ \biggr\} \ ,
\label{atra}
\ea
and from $\tilde{\cal K}$ and $\tilde{\cal A}$, by standard methods, it is
straightforward to obtain the open spectra, encoded in the 
direct-channel amplitudes 
\ba 
{\cal A} &=& \frac{1}{4} \ \biggl\{ \ ( \ V_4 O_4 + O_4 V_4 - S_4 S_4 - C_4 C_4 \ ) 
\ ( \ N^2 P  + D^2 W  \ ) \nonumber \\  
&+&  2 N D \ ( \ O_4S_4 + V_4 C_4 - C_4 O_4 - S_4 V_4 \ ) \ 
{\biggl(\frac{\eta}{\theta_4} \ \biggr)}^2 \nonumber \\ 
&+& ( \ R_N^2 + R_D^2 \ ) \  (\ V_4 O_4 - O_4 V_4 + S_4 S_4 - C_4 C_4 \ ) \ {\biggl( \ \frac{2
\eta}{\theta_2} \ \biggr) \ }^2 \nonumber \\
&+& 2 R_N R_D \ ( \ V_4 C_4 - O_4 S_4 + S_4 V_4 - C_4 O_4 \ ) \ 
{\biggl( \ \frac{\eta}{\theta_3} \ \biggr)}^2  \ \biggr\} 
\label{adir}
\ea
and
\ba
{\cal M} = &- & \frac{1}{4} \ \biggl\{ \  N P \ ( \ \hat{O}_4 \hat{V}_4 + \hat{V}_4 \hat{O}_4 - 
\hat{S}_4 \hat{S}_4 - \hat{C}_4 \hat{C}_4 \ ) \nonumber \\
&-&  D W  \ ( \ \hat{O}_4 \hat{V}_4 + \hat{V}_4 
\hat{O}_4 + \hat{S}_4 \hat{S}_4 + \hat{C}_4 \hat{C}_4 \ ) \nonumber \\ 
&-& N \ ( \ \hat{O}_4 \hat{V}_4 \!-\! \hat{V}_4 \hat{O}_4 \!-\! \hat{S}_4 
\hat{S}_4 \!+\! \hat{C}_4 \hat{C}_4 \ ) \ \left(
\ {2{\hat{\eta}}\over{\hat{\theta}}_2} \ \right)^2  \nonumber \\
&+& D \ ( \ \hat{O}_4 
\hat{V}_4 - \hat{V}_4 \hat{O}_4 + \hat{S}_4 \hat{S}_4
 -  \hat{C}_4 \hat{C}_4 \ ) \ \left( \ 
{2{\hat{\eta}}\over{\hat{\theta}}_2} \ \right)^2  \ \biggr\} \ .
\label{mobdir} 
\ea
The R-R tadpole cancellation conditions require
\ba
N \ &=& \ D \ = \ 32 \quad , \nonumber \\
R_N \ &=& \ R_D \ = \ 0 \quad, 
\label{tad}
\ea
and allow a parametrization in terms of real Chan-Paton multiplicities
of the form
$N=n_1+ n_2$, $D=d_1+ d_2$, $R_N=n_1- n_2$ and $R_D=d_1- d_2$,
with $n_1=n_2=d_1=d_2=16$.  
The massless spectrum can be extracted from
\ba  
{\cal A}_0 + {\cal M}_0 &=& \big[ \ \frac{n_1 (n_1-1)}{2} \ + \   
\frac{n_2(n_2-1)}{2} \ + \  \frac{d_1(d_1+1)}{2} \ + \ \frac{d_2(d_2+1)}{2} \  
\big] \ V_4 O_4  \nonumber \\
 &-& \big[ \ \frac{n_1(n_1-1)}{2} \ + \  
\frac{n_2(n_2-1)}{2} \ + \ \frac{d_1(d_1-1)}{2} \ + \ \frac{d_2(d_2-1)}{2} \ 
\big] \ C_4 C_4  \nonumber \\
&+&  ( \ n_1 d_2 + n_2 d_1 \ ) \ O_4 S_4 - ( \ n_1 d_1 + n_2 d_2 \ ) \ C_4 O_4  \nonumber \\
&+& ( \ n_1 n_2 + d_1 d_2 \ ) \ ( \ O_4 V_4 - S_4 S_4 \ ) \quad ,
\label{spect}
\ea 
and the gauge group is thus $[ SO(16) \times SO(16) ]_{9} \times  [ USp(16)
\times USp(16) ]_{\bar{5}}$.  Supersymmetry, exact in the 9-9 sector, 
where the 
vector multiplets of the two $SO(16)$ are accompanied by a hypermultiplet
in the  ${\bf (16,16,1,1)}$, is broken on
the ${\bar{D}}5$ branes, where the gauge vectors of the two $USp(16)$ are in the
adjoint representation while the left-handed Weyl fermions are again in
reducible antisymmetric representations, now the ${\bf (1,1,120,1)}$ and
the ${\bf (1,1,1,120)}$. 
In addition, there are
four scalars and two right-handed Weyl fermions in the 
${\bf (1,1,16,16)}$, as well as two scalars in the  ${\bf (16,1,1,16)}$, 
two scalars in the  ${\bf (1,16,16,1)}$ and symplectic Majorana-Weyl 
fermions in the  ${\bf (16,1,16,1)}$ and  ${\bf (1,16,1,16)}$.
This model is free of gauge and gravitational 
anomalies and provides an example of type-I vacuum with a stable non-BPS
configuration of BPS branes.  As in the ten-dimensional $USp(32)$ model,
the breaking of supersymmetry yields a tree-level potential
for the NS-NS moduli related to the uncanceled tadpoles, the dilaton and the
internal volume in this case, that reflects the positive tension
resulting from the $O5_-$ planes and the anti-branes.  

Similar considerations apply to generic $Z_N$ orientifolds
in six and four dimensions, as well as to more complicated compactifications,
to wit freely acting orbifolds or non-geometric examples.  
From the low-energy point of view, all  
models with brane supersymmetry breaking exhibit in their spectra
a gauge singlet on the non supersymmetric branes, with the right 
quantum numbers to be a goldstino. As we shall see in the 
next sections, these goldstinos play 
the role of Volkov-Akulov fields that allow consistent couplings
of the gravitinos to the non supersymmetric matter.  Supersymmetry is thus
linearly realized in the bulk and on some branes, while it is non-linearly 
realized on other (anti)branes. 

\section{Low-energy couplings for the Sugimoto model}
This section builds on \cite{dm2}, where 
the low-energy effective action for the $USp(32)$ model was 
constructed, to lowest order in the fermi fields, requiring that supersymmetry 
be non-linearly realized on the ${\bar{D}}9$ branes, and thus obtaining 
consistent couplings for the gravitino. Our aim is to show  
how {\it all} the couplings of \cite{dm2} can be written in a geometric form. 

Let us start by reviewing the work of \cite{dm2}, in a slightly different 
notation, working in the Einstein frame with metric signature
$(+,-,...,-)$. As we have seen in the previous section, the Sugimoto model 
results from a different IIB orientifold projection
with respect to the one leading to the type-I $SO(32)$ theory. The closed 
spectrum is identical in the two cases, and comprises at the massless
level the (1,0) supergravity multiplet, with the vielbein
$e_\m{}^m$, a 2-form $B_{\m\n}$, the dilaton $\phi$, 
a left-handed gravitino $\psi_\m$ and a right-handed dilatino $\chi$, while 
the open sector describes a $USp(32)$ gauge group, whose gauge
boson $A_\m$ is accompanied by a massless Majorana-Weyl spinor 
in the {\it reducible} antisymmetric representation, that contains a 
spinor $\l$ in the {\bf 495} and a
singlet $\th$.   
Dudas and Mourad identified in \cite{dm2} this singlet as the goldstino of 
supersymmetry, that is non-linearly realized on the brane modes, 
consistently with its breaking at the string scale. 
Starting from this, they constructed the low-energy effective action for 
the Sugimoto model up to quartic terms in the spinors.

Let us briefly review how a single spinor can be treated {\it \`a la} 
Volkov-Akulov \cite{samwess} as a goldstino of global supersymmetry. Let us 
restrict our attention to the ten dimensional case, considering a 
Majorana-Weyl fermion $\th$ with the 
supersymmetry transformation
\be
\d \th =\e  - \frac{i}{2}(\bar{\e} \g^\m \th ) \de_\m \th \label{va}.
\ee
The commutator of two such transformations is a translation,
\be
[\d_1 , \d_2 ] \th = -i (\bar{\e}_2 \g^\m \e_1 )  \de_\m \th \quad ,
\ee
and thus eq. (\ref{va}) provides a realization of supersymmetry.
In order to write a 
Lagrangian for $\th$ invariant under eq. (\ref{va}), let us define
the 1-form
\be
e_\m{}^m= \d_\m^m -\frac{i}{2}(\bar{\th} \g^m \de_\m \th )\quad ,
\ee
whose supersymmetry transformation is 
\be
\d e^m = -L_\xi e^m \quad ,
\ee
with $L_\xi$ the Lie derivative with respect to
\be
\xi_\m= -\frac{i}{2}(\bar{\th}\g_\m \e ) \quad . \label{xi}
\ee
The action of supersymmetry on $e$ is thus a general coordinate 
transformation, with a parameter depending on $\th$, and therefore
\be
{\cal L} = -\det e 
\ee
is clearly an invariant Lagrangian. 
Expanding the determinant, one can see that the energy has a positive 
vacuum expectation value, and supersymmetry is thus spontaneously broken.
Using the same technique, for a generic field $A$ that transforms 
under supersymmetry as 
\be
\d A = - L_\xi A \quad , \label{susygct}
\ee
defining the induced metric as $g_{\m\n}=e_\m{}^m e_{\n m}$, a 
supersymmetric Lagrangian in flat space is determined by
the substitution
\be
{\cal L}(\eta,A) \rightarrow e{\cal L}(g,A) \quad .
\ee

We can now review the results of \cite{dm2}, and to this end
we begin by considering the Lagrangian for the closed sector
\bea
e^{-1} {\cal{L}}_{closed}  =
&-& \frac{1}{4} R +\frac{1}{2}\de_\m \phi \de^\m \phi +\frac{1}{6}
e^{-2\phi}H_{\m\n\r} H^{\m\n\r} \nonumber\\
&-&\frac{i}{2} (\bar{\psi}_\m \g^{\m\n\r} D_\n \psi_\r )+\frac{i}{2}(\bar{\chi}
\g^\m D_\m \chi )+\frac{1}{\sqrt{2}}(\bar{\psi}_\m \g^\n \g^\m \chi )\de_\n \phi 
\nonumber\\
&-& \frac{i}{12\sqrt{2}} e^{-\phi} H_{\m\n\r} (\bar{\psi}_\s \g^{\s\d\m\n\r}\psi_\d
)+\frac{i}{2\sqrt{2}}e^{-\phi} H_{\m\n\r} (\bar{\psi}^\m \g^\n \psi^\r)\nonumber \\
&+& \frac{1}{12}
e^{-\phi}H_{\m\n\r} (\bar{\psi}_\s \g^{\m\n\r} \g^\s \chi) \quad ,
\label{sugra}
\eea
that provides a linear realization of the minimal (1,0) ten-dimensional
supersymmetry, and is thus invariant under the local 
supersymmetry transformations
\bea
& & \d e_\m{}^m = -i (\bar{\e} \g^m \psi_\m )\quad , \nonumber\\
& & \d B_{\m\n}= -\frac{i}{\sqrt{2}}e^\phi (\bar{\e}\g_{[\m}\psi_{\n]} )-\frac{1}{4}
e^\phi (\bar{\e}\g_{\m\n} \chi )\quad, \nonumber \\
& & \delta \phi =-\frac{1}{\sqrt{2}} (\bar{\e} \chi )\quad ,\nonumber\\
& & \d \psi_\m =D_\m \e +\frac{1}{24\sqrt{2}}e^{-\phi}H^{\n\r\s}\g_{\m\n\r\s} \e
-\frac{3}{8\sqrt{2}}e^{-\phi} H_{\m\n\r} \g^{\n\r}\e \quad , \nonumber\\
& & \d \chi =-\frac{i}{\sqrt{2}}\de_\m \phi \g^\m \e-\frac{i}{12}e^{-\phi}
H^{\m\n\r} \g_{\m\n\r} \e \quad . 
\eea

In the supersymmetric case,
when these bulk modes are coupled to a gauge multiplet supported on the 9-branes and
containing a vector $A_\m$ and a left-handed gaugino $\l$ both in the adjoint representation 
of $SO(32)$,
supersymmetry requires that the 3-form 
$H_{\m\n\r}$ include a Chern-Simons coupling, so that 
\be
H_{\m\n\r}= 3 \de_{[\m} B_{\n\r ]} +\sqrt{2} \w_{\m\n\r} \quad ,\label{3form}
\ee
where $\w_{\m\n\r}$ is the Chern-Simons 3-form defined as 
\be
\w = A dA -\frac{2i}{3} A^3 \quad ,
\ee
and this leads to the modified Bianchi identity
\be
\de_{[\m }H_{\n\r\s ]} =\frac{3}{\sqrt{2}}tr (F_{[\m\n} F_{\r\s ]} )\quad .
\label{bianchi}
\ee
The Lagrangian for supergravity coupled to vector multiplets is then \cite{bddv}
\bea
e^{-1} {\cal{L}} = 
&-& \frac{1}{4} R +\frac{1}{2}\de_\m \phi \de^\m \phi +\frac{1}{6}
e^{-2\phi}H_{\m\n\r} H^{\m\n\r} -\frac{1}{2}e^{-\phi}tr (F_{\m\n}F^{\m\n} )
\nonumber\\
&-& \frac{i}{2} (\bar{\psi}_\m \g^{\m\n\r} D_\n \psi_\r )+\frac{i}{2}(\bar{\chi}
\g^\m D_\m \chi )+\frac{1}{\sqrt{2}}(\bar{\psi}_\m \g^\n \g^\m \chi )\de_\n \phi 
\nonumber\\
&-& \frac{i}{12\sqrt{2}} e^{-\phi} H_{\m\n\r} (\bar{\psi}_\s \g^{\s\d\m\n\r}\psi_\d
)+\frac{i}{2\sqrt{2}}e^{-\phi} H_{\m\n\r} (\bar{\psi}^\m \g^\n \psi^\r)\nonumber \\
&+& \frac{1}{12}
e^{-\phi}H_{\m\n\r} (\bar{\psi}_\s \g^{\m\n\r} \g^\s \chi) +i tr (\bar{\l}\g^\m
D_\m \l )\nonumber \\
&-& \frac{1}{2} e^{-\frac{1}{2}\phi } tr [F^{\m\n} (\bar{\l} \g_{\m\n} \chi)]
+\frac{i}{\sqrt{2}}e^{-\frac{1}{2}\phi }tr [F^{\m\n} (\bar{\l} \g_\r \g_{\m\n}
\psi^\r )]\nonumber\\
&-& \frac{i}{6\sqrt{2}}e^{-\phi }H^{\m\n\r}tr (\bar{\l} \g_{\m\n\r} \l ) \ , 
\label{lag1}
\eea
up to quartic terms in the fermions.
The supersymmetry transformations of the bulk fields
$e_\m{}^m$, $\phi$, $\psi_\m$ and $\chi$ are as before, while 
for the gauge multiplet
\bea
& & \d A_\m =-\frac{i}{\sqrt{2}} e^{\frac{1}{2}\phi}(\bar{\e} \g_\m \l ) \quad ,
\nonumber\\
& & \d \l =-\frac{1}{2\sqrt{2}}  e^{-\frac{1}{2}\phi}F^{\m\n}\g_{\m\n}\e \quad . 
\label{susy1}
\eea
Gauge invariance of $H$ requires that under vector gauge
transformations $B$ transform as 
\be
\d B = -\sqrt{2} tr (\L dA ) \ ,
\ee
and in order that gauge and 
supersymmetry transformations commute, up to a tensor gauge transformation,
one has to add a term to the supersymmetry variation of $B_{\m\n}$, obtaining
\be
\d B_{\m\n}= -\frac{i}{\sqrt{2}}e^\phi (\bar{\e}\g_{[\m}\psi_{\n]} )-\frac{1}{4}
e^\phi (\bar{\e}\g_{\m\n} \chi ) +2\sqrt{2} tr (A_{[\m } \d A_{\n ]})
\quad. \label{B2}
\ee

In order to couple the Lagrangian (\ref{sugra})
to non-supersymmetric matter, one must construct from the fields of the 
supergravity multiplet quantities whose supersymmetry variations are
general coordinate transformations with the parameter $\xi_\m$ of
eq. (\ref{xi}). 
We thus define
\be
\hat{\phi}=\phi +\frac{1}{\sqrt{2}}(\bar{\th}\chi) +\frac{i}{24\sqrt{2}}e^{-\phi}
(\bar{\th} \g_{\m\n\r} \th ) H^{\m\n\r} \label{phihat}
\ee
so that
\be
\d \hat{\phi} = -\xi^{\m} \de_\m \hat{\phi} =\d_{gct}\hat{\phi}
\ee
and
\bea
\hat{e}_\m{}^m &=& e_\m{}^m +i (\bar{\th}\g^m \psi_\m )-\frac{i}{2}(\bar{\th}\g^m 
D_\m \th )-\frac{i}{48\sqrt{2}}e_\m{}^m e^{-\phi} (\bar{\th}\g_{\n\r\s}\th )
H^{\n\r\s}
\nonumber\\
&+& \frac{i}{16\sqrt{2}}e^{-\phi} (\bar{\th} \g_{\m\n\r} \th )H^{m \n\r}+
\frac{3i}{16\sqrt{2}} e^{-\phi} (\bar{\th} \g^{m \n\r} \th )H_{\m\n\r} 
\label{ehat}
\eea
so that
\be
\d \hat{e}_\m{}^m = \d_{gct} \hat{e}_\m{}^m +\L^m{}_n \hat{e}_\m{}^n \quad ,
\ee
where the parameter of the local Lorentz transformation is
\be
\L^{mn} =\frac{i}{2}(\bar{\th} \g^\r \e )\w_\r{}^{mn} 
+\frac{i}{24\sqrt{2}} e^{-\phi}
(\bar{\th} \g^{mn\n\r\s}\e )H_{\n\r\s} +\frac{3i}{4\sqrt{2}} e^{-\phi}
(\bar{\th} \g_\r \e )H^{\r mn} \quad .
\ee 
In constructing a Lagrangian invariant under non-linear 
supersymmetry that couples supergravity to non-supersymmetric matter,
it is important to notice that 
eq. (\ref{3form}) still holds, because of anomaly cancellation. For the same 
reason, the variation of $B$ is still given by eq. (\ref{B2}), once one 
uses the new transformation for $A_\m$, 
\be
\d A_\m = F_{\m\n} \xi^\n \quad .
\ee
Observe that this covariant expression for $\d A_\m$ contains the proper 
coordinate transformation, together with
an additional gauge transformation of parameter
\be
\L = \xi^\m A_\m \quad .
\ee
The supersymmetry transformation of the spinor $\l$ in the {\bf 495} of $USp(32)$ 
will not be taken 
into account in this discussion, since it contains higher-order fermi terms.
One can now include \cite{dm2} the kinetic term for $A_\m$ and the 
dilaton tadpole in a Lagrangian that is supersymmetric up to terms  
quartic in the fermions, considering 
\bea
{\cal{L}}  &=& {\cal{L}}_{closed} - \frac{1}{2} \hat{e} e^{-\hat{\phi}}
\hat{g}^{\m\r} \hat{g}^{\n\s}tr( F_{\m\n} F_{\r\s} ) -\L \hat{e}
e^{\frac{3}{2}\hat{\phi}} \nonumber \\
&+& ie \, tr (\bar{\l} \g^\m D_\m \l ) 
-\frac{ie}{6\sqrt{2}}e^{-\phi}H^{\m\n\r} 
\, tr(\bar{\l} \g_{\m\n\r} \l ) \quad ,
\label{geom}
\eea
where in the Sugimoto model $\L = 64 T_9$, with $T_9$ the anti-brane tension. 
By string considerations, one can show that the coefficient of the coupling 
of $H$ to $\l^2$, not constrained by supersymmetry at this level, is the 
same as in the supersymmetric case \cite{dm2}. 
Actually, the Lagrangian of eq. (\ref{geom}) is still 
not invariant under supersymmetry, since the 
inclusion of the Chern-Simons term and the consequent
modification of the Bianchi identity for $H$ generate contributions 
proportional to $F \wedge F$ in the variation of ${\cal{L}}_{closed}$. 
Up to higher order fermionic terms, however, these are exactly canceled by 
the variation of the additional terms
\bea
& & \frac{1}{6 !\sqrt{2}}\e^{\m_1 ...\m_6 \m\n\r\s}
[\frac{3i}{\sqrt{2}} e^{-\phi} (\bar{\th} \g_{\m_1 ...\m_5} \psi_{\m_6})
-\frac{1}{4}  e^{-\phi}(\bar{\th} \g_{\m_1 ...\m_6} \chi )\nonumber\\ 
& & +\frac{i}{8\sqrt{2}}
e^{-2\phi} \de_\t \phi (\bar{\th} \g_{\m_1 ...\m_6}{}^\t \th ) 
-\frac{3i}{2\sqrt{2}} e^{-\phi} (\bar{\th} \g_{\m_1 ...\m_5}D_{\m_6}\th )
\nonumber \\
& & -\frac{3i}{8} e^{-2\phi} (\bar{\th} \g_{\m_1 ...\m_5 \t\d} \th )
H_{\m_6}{}^{\t\d} -\frac{5i}{2}e^{-2\phi} 
(\bar{\th} \g_{\m_1 \m_2 \m_3} \th )
H_{\m_4 \m_5 \m_6} ] tr(F_{\m\n} F_{\r\s} ) \quad .
\label{nongeom}
\eea
To summarize,  up to quartic fermionic terms the Lagrangian is 
\bea
{\cal{L}}  &=& {\cal{L}}_{closed} - \frac{1}{2} \hat{e} e^{-\hat{\phi}}
\hat{g}^{\m\r} \hat{g}^{\n\s}tr( F_{\m\n} F_{\r\s} ) -\L \hat{e}
e^{\frac{3}{2}\hat{\phi}} \nonumber \\
&+& ie \, 
tr (\bar{\l} \g^\m D_\m \l ) -\frac{ie}{6\sqrt{2}}e^{-\phi}H^{\m\n\r}
\, tr (\bar{\l} \g_{\m\n\r} \l ) \nonumber\\
&+&  \, \frac{1}{6 !\sqrt{2}}\e^{\m_1 ...\m_6 \m\n\r\s}
[\frac{3i}{\sqrt{2}} e^{-\phi} (\bar{\th} \g_{\m_1 ...\m_5} \psi_{\m_6})
-\frac{1}{4}  e^{-\phi}(\bar{\th} \g_{\m_1 ...\m_6} \chi )\nonumber\\ 
&+& \frac{i}{8\sqrt{2}}
e^{-2\phi} \de_\t \phi (\bar{\th} \g_{\m_1 ...\m_6}{}^\t \th ) 
-\frac{3i}{2\sqrt{2}} e^{-\phi} (\bar{\th} \g_{\m_1 ...\m_5}D_{\m_6}\th )
\nonumber \\
&-& \frac{3i}{8} e^{-2\phi} (\bar{\th} \g_{\m_1 ...\m_5 \t\d} \th )
H_{\m_6}{}^{\t\d} -\frac{5i}{2}e^{-2\phi} 
(\bar{\th} \g_{\m_1 \m_2 \m_3} \th )
H_{\m_4 \m_5 \m_6} ] tr(F_{\m\n} F_{\r\s} ) \quad .\label{lagB2}
\eea
As noticed in \cite{dm2}, in this formulation 
a geometric description for the terms in eq. (\ref{nongeom}) is  
not possible, 
{\it i.e.} it is not possible to rewrite them in terms of properly
dressed bulk 
fields adding fermionic bilinears containing the goldstino. 
We can now explain why this is the case,
and moreover we can also show how a geometric description is possible, 
after performing a duality transformation 
to a 6-form gauge field. 

Let us again begin with standard results: performing a duality transformation 
on eq. (\ref{lag1}), one  obtains a new Lagrangian,
with a 6-form rather than a 2-form, coupled
to vector multiplets \cite{chamseddine}.
Technically, this is performed starting from the first-order Lagrangian
\bea
{\cal L} &=& 
\frac{1}{6} e^{-2\phi} H_{\m\n\r}H^{\m\n\r} +\frac{1}{3 \cdot 6!}
\e^{\m_1 ...\m_7 \m\n\r } \de_{\m_1} \tilde{B}_{\m_2 ...\m_7} H_{\m\n\r} 
\nonumber\\
&+& \frac{1}{6! \sqrt{2}}\e^{\m_1 ...\m_6 \m\n\r\s }\tilde{B}_{\m_1 ...\m_6}
tr (F_{\m\n}
F_{\r\s})
\label{lduality}
\eea
that contains both the 2-form and the 6-form.
The field equation for $\tilde{B}_6$ is then exactly the Bianchi 
identity of eq. (\ref{bianchi})
for $H_3$, while the field equation for $H_3$ is
\be
e^{-\phi} H_3 = e^\phi * \tilde{H}_7 \quad ,
\label{dual}
\ee
where $\tilde{H}_7 =d \tilde{B}_6$.
The Lagrangian obtained substituting this relation in (\ref{lduality}) and
redefining $\tilde{H}_7 \rightarrow H_7$,
\be
{\cal L}=\frac{1}{7 !}e^{2\phi}H_{\m_1 ...\m_7 }H^{\m_1 ...\m_7 }+
\frac{1}{6! \sqrt{2}}\e^{\m_1 ...\m_6 \m\n\r\s }B_{\m_1 ...\m_6}tr (F_{\m\n}
F_{\r\s}) \quad ,
\ee
shows how the Chern-Simons term in $H_3$ is replaced by the 
Wess-Zumino term $B \wedge F \wedge F$.

If one performs this duality transformation in (\ref{lag1}),
one ends up with the Lagrangian originally 
obtained by Chamseddine \cite{chamseddine}:
\bea
e^{-1} {\cal{L}} =  
&-&\frac{1}{4} R +\frac{1}{2}\de_\m \phi \de^\m \phi +\frac{1}{7!}
e^{2\phi}H_{\m_1 ...\m_7} H^{\m_1 ...\m_7} 
-\frac{1}{2}e^{-\phi}tr (F_{\m\n}F^{\m\n} )
\nonumber\\
&+& \frac{1}{6! \sqrt{2}}\e^{\m_1 ...\m_6 \m\n\r\s } B_{\m_1 ...\m_6}
tr (F_{\m\n}
F_{\r\s}) -\frac{i}{2} (\bar{\psi}_\m \g^{\m\n\r} D_\n \psi_\r ) \nonumber \\
&+&\frac{i}{2}(\bar{\chi}
\g^\m D_\m \chi )+\frac{1}{\sqrt{2}}(\bar{\psi}_\m \g^\n \g^\m \chi )\de_\n \phi 
+\frac{i}{240\sqrt{2}} e^{\phi} H^{\m_1 ...\m_7} 
(\bar{\psi}_{\m_1} \g_{\m_2 ...\m_6} \psi_{\m_7} ) \nonumber \\
&+& \frac{i}{2 \cdot 7!\sqrt{2}}e^{\phi} 
H^{\m_1 ...\m_7} (\bar{\psi}^\m \g_{\m\m_1 ... \m_7 \n} \psi^\n) -\frac{1}{2 \cdot 7!}
e^{\phi}H^{\m_1 ...\m_7} (\bar{\psi}^\s \g_{\m_1 ...\m_7} \g_\s \chi) \nonumber \\
&+& i tr (\bar{\l}\g^\m
D_\m \l )-\frac{1}{2} e^{-\frac{1}{2}\phi } tr [F^{\m\n} (\bar{\l} \g_{\m\n} \chi)] 
\nonumber \\
&+& \frac{i}{\sqrt{2}}e^{-\frac{1}{2}\phi }tr [F^{\m\n} (\bar{\l} \g_\r \g_{\m\n}
\psi^\r )]+\frac{i}{7!\sqrt{2}}e^{\phi }H^{\m_1 ...\m_7}tr(\bar{\l} \g_{\m_1 ...\m_7} \l ) 
\quad .
\label{lag2}
\eea
The corresponding supersymmetry transformations are obtained 
from eq. (\ref{susy1}) performing the redefinition of eq. (\ref{dual}) 
on the variations of $\psi_\m$ and $\chi$, leaving the variations of 
$e_\m{}^m$, $\phi$, $A_\m$ and $\l$ unaffected
and replacing the variation of the 2-form with 
\be
\d B_{\m_1 ...\m_6}= -\frac{3i}{\sqrt{2}}e^{-\phi} (\bar{\e}\g_{[\m_1 ...\m_5}
\psi_{\m_6 ]})+\frac{1}{4} e^{-\phi} (\bar{\e}\g_{\m_1 ...\m_6 }\chi )\quad .
\label{B6}
\ee
Notice that the supersymmetry variation of the 6-form does not include a term 
depending on the vector field. This reflects the  
fact that the 6-form is inert under gauge transformations, 
since its field-strength does not contain a Chern-Simons form, that
in this case would enter higher-derivative couplings not present in the
effective supergravity.

One can now couple the supergravity multiplet expressed in terms of the
6-form to non-supersymmetric matter. In order to do this, 
together with $\hat{\phi}$ and $\hat{e}_\m{}^m$ of eqs. (\ref{phihat}) and 
(\ref{ehat}), one must define the dressed 6-form
\bea
\hat{B}_{\m_1 ...\m_6}  &=& B_{\m_1 ...\m_6} + 
\frac{3i}{\sqrt{2}} e^{-\phi} (\bar{\th} \g_{[\m_1 ...\m_5} \psi_{\m_6 ]})
-\frac{1}{4}  e^{-\phi}(\bar{\th} \g_{\m_1 ...\m_6} \chi )\nonumber\\ 
&+& \frac{i}{8\sqrt{2}}
e^{-2\phi} \de_\t \phi (\bar{\th} \g_{\m_1 ...\m_6}{}^\t \th ) 
-\frac{3i}{2\sqrt{2}} e^{-\phi} (\bar{\th} \g_{[ \m_1 ...\m_5}D_{\m_6 ]}\th )
\nonumber \\
&-& \frac{3i}{8} e^{-2\phi} (\bar{\th} \g_{[ \m_1 ...\m_5 \t\d} \th )
H_{\m_6 ]}{}^{\t\d} -\frac{5i}{2}e^{-2\phi} 
(\bar{\th} \g_{[ \m_1 \m_2 \m_3} \th )
H_{\m_4 \m_5 \m_6 ]}\quad , \label{B6hat}
\eea
whose supersymmetry transformation is a coordinate 
transformation, up to
an additional tensor gauge transformation of parameter
\be
\L_{\m_1 ... \m_5 } =-\frac{i}{4 \sqrt{2}}e^{-\phi} (\bar{\th}\g_{\m_1 ... \m_5 }\e ) \quad.
\ee
We have intentionally written the last line of eq. (\ref{B6hat}) in terms of the dual
3-form, using eq. (\ref{dual}), so that the similarity 
with eq. (\ref{nongeom}) be more transparent. 
The Lagrangian for the closed sector,
\bea
e^{-1} \tilde{\cal{L}}_{closed}  =
&-& \frac{1}{4} R +\frac{1}{2}\de_\m \phi \de^\m \phi +\frac{1}{6}
e^{2\phi}H_{\m_1 ...\m_7} H^{\m_1 ...\m_7} \nonumber\\
&-& \frac{i}{2} (\bar{\psi}_\m \g^{\m\n\r} D_\n \psi_\r )+\frac{i}{2}(\bar{\chi}
\g^\m D_\m \chi )+\frac{1}{\sqrt{2}}(\bar{\psi}_\m \g^\n \g^\m \chi )\de_\n \phi 
\nonumber\\
&+& \frac{i}{240\sqrt{2}} e^{\phi} H^{\m_1 ...\m_7} 
(\bar{\psi}_{\m_1} \g_{\m_2 ...\m_6}\psi_{\m_7}
)+\frac{i}{2\sqrt{2}7!}e^{\phi} 
H^{\m_1 ...\m_7} (\bar{\psi}^\m \g_{\m \m_1 ...\m_7 \n} \psi^\n)\nonumber \\
&+& \frac{1}{2\cdot 7!}
e^{\phi}H^{\m_1 ...\m_7} (\bar{\psi}^\s \g_{\m_1 ...\m_7} \g_\s \chi) 
\label{sugra2}
\eea
is simply obtained performing the duality 
transformation in eq. (\ref{sugra}), while
the same duality in eq. 
(\ref{lagB2}) gives 
\bea
{\cal{L}}  &=& \tilde{\cal{L}}_{closed} 
 - \frac{1}{2} \hat{e} e^{-\hat{\phi}}
\hat{g}^{\m\r} \hat{g}^{\n\s}tr( F_{\m\n} F_{\r\s} ) -\L \hat{e}
e^{\frac{3}{2}\hat{\phi}} \nonumber \\
&+& ie tr (\bar{\l} \g^\m D_\m \l ) 
+\frac{i e}{7!\sqrt{2}}
e^{\phi }H^{\m_1 ...\m_7}tr(\bar{\l} \g_{\m_1 ...\m_7} \l ) \nonumber\\
&+& \frac{1}{6!\sqrt{2}}\e^{\m_1 ...\m_6 \m\n\r\s} \hat{B}_{\m_1 ... \m_6 }
tr (F_{\m\n} F_{\r\s} ) \quad .\label{lagB6}
\eea
Note that in this Lagrangian {\it all} terms containing the goldstino are
grouped in redefinitions of the bulk fields, and therefore 
all couplings are written in a geometric form.
This result concludes this section: for the
ten-dimensional $USp(32)$ model a fully geometric description is possible if
one formulates it in terms of the 6-form, since in this case the 
Chern-Simons term is higher derivative, and thus is not in the 
low-energy effective action. More precisely, as we have seen, 
duality maps the Chern-Simons term 
into the Wess-Zumino term, and this falls simply into a geometric
form.  The result is still valid in presence of additional
brane-antibrane pairs, since
the introduction of supersymmetric vectors does not modify the 
field strength relative to the 6-form potential in the low-energy effective action.
In the dual theory, although the field strength of the 2-form is modified, no additional
terms containing the goldstino have to be added to the low-energy lagangian. 
As we shall see, this feature is common to the six-dimensional case. 

\section{Geometric couplings in six-dimensional models}
In this section we construct the low-energy couplings for 
six-dimensional type-I models with brane supersymmetry breaking 
\cite{bsb}. 
As explained in  Section 2, all the features 
of brane supersymmetry breaking are present in the 
$T^4 / Z_2$ orientifold of \cite{bsb}, where a change of the 
orientifold projection leads to $D9$ branes and $\bar{D}5$ branes. 
The spectrum has $(1,0)$ supersymmetry in the closed 
and 9-9 sectors, while supersymmetry is broken in the 9-$\bar{5}$ and
$\bar{5}$-$\bar{5}$ sectors. The gauge group is $SO(16) \times SO(16)$ on 
the $D9$ branes and $USp(16) \times USp(16)$ on the ${\bar{D}}5$ branes, 
if all the ${\bar{D}}5$ branes are at a fixed point.

One of the peculiar features of low-energy effective actions for six-dimensional
type-I models with minimal supersymmetry is the fact that they embody reducible gauge and 
supersymmetry anomalies, to be canceled by fermion loops. Consequently, the 
Lagrangian is determined imposing the closure of the Wess-Zumino consistency 
conditions, rather than
by the requirement of supersymmetry.
The coupling of $(1,0)$ supergravity to tensor multiplets 
was obtained in \cite{romans} to lowest order in the fermi fields, while \cite{nsez1} 
considered a single tensor multiplet coupled to  vector multiplets and
hypermultiplets to all orders in the fermi fields, without taking into account 
the anomalous couplings. The coupling of vector multiplets to an arbitrary number of  
tensor multiplets 
was obtained to lowest order in the fermi fields in \cite{ggs} in the covariant 
formulation and in \cite{ggs2} in the consistent formulation, 
and these results have been generalized in \cite{ggs3,rs2} to all orders in 
the fermi fields. Additional couplings to hypermultiplets have been included in
\cite{nsez2}, and in \cite{fabio} the complete coupling of supergravity 
to vector, tensor and hypermultiplets
has been obtained. 
In order to describe the supersymmetric part of the model of \cite{bsb}, let
us briefly review these results, taking into account only the 
lowest order fermi terms. 
 
We first summarize the field content of the multiplets. The gravitational 
multiplet contains the vielbein $e_\m{}^m$, a 2-form and 
a left-handed gravitino $\psi_\m^A$, the tensor multiplet contains 
a 2-form, a scalar and a right-handed tensorino, the  
vector multiplet from the 9-9 sector contains a vector $A^{(9)}_\m$ and a left-handed gaugino 
$\l^{(9)A}$, and finally the 
hypermultiplet contains four scalars and a right-handed hyperino. 
In the presence of $n_T$ tensor multiplets,
the tensorinos are denoted by 
$\chi^{MA}$, where $M=1,...,n_T$ is an $SO(n_T )$ index. The index $A=1,2$
is in the fundamental representation of $USp(2)$, and the gravitino, the 
tensorinos and the gauginos are $USp(2)$ doublets satisfying 
the symplectic-Majorana condition
\be
\psi^A =\e^{AB} C \bar{\psi}^T_B \quad . 
\ee
The $n_T$ scalars in the tensor multiplets 
parametrize the coset 
$SO(1,n_T)/SO(n_T)$, while the $(n_T +1)$ 2-forms from the gravitational 
and tensor multiplets are collectively 
denoted by $B^r_{\m\n}$,
with $r=0,...,n_T$ valued in the fundamental representation of $SO(1,n_T)$, 
and their field-strengths satisfy 
(anti)self-duality conditions.
Finally, taking into account $n_H$ hypermultiplets,
the hyperinos are denoted by $\Psi^{a}$, where $a=1,...,2n_H$ is a 
$USp(2n_H )$ index, and the symplectic-Majorana condition for these spinors
is
\be
\Psi^a =\W^{ab} C \bar{\Psi}^T_b \quad,
\ee
where $\W^{ab}$ is the 
antisymmetric invariant tensor of $USp(2n_H)$ (see ref. \cite{fabio} for more details). 

The scalars  $\Phi^{\bar{\a}}$ $(\bar{\a}=1,...,n_T )$ in the tensor multiplets
can be described, following \cite{romans}, in terms of 
the $SO(1,n_T)$ matrix
\be
V = \pmatrix{v_r \cr x^M_r} \quad ,
\ee
whose elements are functions of $\Phi^{\bar{\a}}$ satisfying  the constraints
\be
v^r v_r =1 \quad , \qquad  v_r v_s - x^M_r x^M_s = \eta_{rs}\quad , \qquad 
v^r x^M_r =0 \quad . \label{scalars}
\ee
The vielbein $V_{\bar{\a}}^M$ of the internal manifold 
is related to $v^r$ and $x^{Mr}$ by 
\be
V_{\bar{\a}}^M =v^r \de_{\bar{\a}} x^M_r \quad ,
\ee
where $\de_{\bar{\a}} = \de / \de \Phi^{\bar{\a}}$.
The metric of the internal manifold is $g_{\bar{\a}\bar{\b}} = V_{\bar{\a}}^M
V_{\bar{\b}}^M$.

The hyper-scalars $\phi^\a$  $(\a=1,...,4n_H)$ are coordinates of a quaternionic
manifold, that is a manifold whose holonomy group is contained in 
$USp(2) \times USp(2n_H)$.
In the model of \cite{bsb}, the hypermultiplets in the closed sector are neutral,
while the hypermultiplets in the 9-9 sector are charged under the gauge group
$SO(16) \times SO(16)$, that  corresponds
to an isometry of the manifold parametrized by the hyper-scalars.  
We denote by $V_\a^{aA}(\phi )$ the vielbein of the quaternionic manifold, 
where the index structure corresponds to the requirement that the 
holonomy be contained
in $USp(2)\times USp(2n_H )$. The internal $USp(2)$ and
$USp(2n_H)$ connections are then denoted, respectively, by 
${\cal{A}}_\a^A{}_B$  and ${\cal{A}}_\a^a{}_b$, that in our conventions 
are anti-hermitian matrices. The index $\a =1,...,4 n_H$ is a curved 
index on the quaternionic manifold, while
the field-strengths of the connections are
\bea
& & {\cal{F}}_{\a\b}{}^A{}_B = \de_{\a} {\cal{A}}_\b^A{}_B -\de_{\b} 
{\cal{A}}_\a^A{}_B
+[ {\cal{A}}_\a , {\cal{A}}_\b ]^A{}_B \quad ,\nonumber\\
& & {\cal{F}}_{\a\b}{}^a{}_b = \de_{\a} {\cal{A}}_\b^a{}_b -\de_{\b} 
{\cal{A}}_\a^a{}_b
+[ {\cal{A}}_\a , {\cal{A}}_\b ]^a{}_b \quad ,
\eea
where $\de_\a = \de / \de \phi^\a $. 
The condition that the vielbein $V_\a^{aA}(\phi )$ 
be covariantly constant gives the relations \cite{bw}
\bea
& & V^{\a}_{aA} V^{\b}_{bB}g_{\a\b}=\W_{ab} \e_{AB} \quad ,\nonumber\\
& & V^{\a}_{aA} V^{\b bA} + V^{\b}_{aA} V^{\a bA} =\frac{1}{n_{H}}
g^{\a\b}\delta^b_a 
\quad , \nonumber\\
& & V^{\a}_{aA} V^{\b aB} +V^{\b}_{aA} V^{\a aB}=g^{\a\b}\delta^B_A \quad ,
\eea 
where $\W_{ab}$ is the antisymmetric invariant tensor of $USp(2n_H)$. 
The field-strength of the $USp(2)$ connection ${\cal{A}}_\a^A{}_B$  
is naturally constructed in terms of $V_{\a}^{aA}$ by the relation 
\be
{\cal{F}}_{\a\b AB}= V_{\a aA}V_\b^a{}_B +V_{\a aB}V_\b^a{}_A \quad ,
\ee
and then the cyclic identity for the internal curvature tensor implies 
that the field-strength of
the $USp(2n_H)$ connection ${\cal{A}}_\a^a{}_b$ has the form
\be
{\cal{F}}_{\a\b ab}= V_{\a aA}V_{\b b}{}^A +V_{\a bA}V_{\b a}{}^A  + \W_{abcd}
V_\a^{dA}V_\b^c{}_A \quad , 
\ee 
where $\W_{abcd}$ is totally symmetric in its indices \cite{bw}.

Denoting with  $A_\m^{(9)i}$ the gauge fields under which the hypermultiplets are
charged (the index $i$ runs in the adjoint of the gauge group), 
under the gauge transformations 
\be
\delta A_\m^{(9)i} = D_\m \L^{(9)i}  
\ee
the scalars transform as
\be
\delta \phi^\a = \L^{(9)i} \xi^{\a i} \quad ,
\ee
where $\xi^{\a i}$ are the Killing vectors corresponding to
the isometry that we are gauging.
The covariant derivative for the scalars is then
\be
D_\m \phi^\a =\de_\m \phi^\a - A_\m^{(9)i} \xi^{\a i} \quad .
\ee
One can correspondingly define the covariant derivatives for the spinors in 
a natural way, adding the composite connections
$D_\m \phi^\a {\cal{A}}_\a$. 
For instance, the covariant derivative for the hyperinos 
$\Psi^a$ will contain the connections $D_\m \phi^\a {\cal{A}}_\a^a{}_b$, 
while the covariant derivative for the gravitino and the tensorinos 
will contain the connections
$D_\m \phi^\a {\cal{A}}_\a^A{}_B$. The covariant derivatives for the 
gauginos $\l^{(9)iA}$ are 
\be
D_\m \l^{(9)iA}=\de_\m \l^{(9)iA} +\frac{1}{4}\w_{\m mn}\g^{mn}\l^{(9)iA}
+ D_\m \phi^\a {\cal{A}}_\a^A{}_B \l^{(9)i B}+f^{ijk} A_\m^{(9)j} \l^{(9)kA}
\quad ,
\ee
where $f^{ijk}$ are the structure constants of the group.

We use the method of Pasti, Sorokin and Tonin (PST) \cite{pst} 
in order to write a covariant action for 
fields that satisfy self-duality conditions. 
For a self-dual 3-form in six dimensions the PST action 
\be
{\cal L}_{PST} = \frac{1}{12}H_{\m\n\r} H^{\m\n\r} -
\frac{1}{4}\frac{\de^\m \Xi \de^\s \Xi}{(\de \Xi )^2 } H^-_{\m\n\r} 
H^-_\s{}^{\n\r}\quad ,
\ee
where $H^- = H- *H$ and $\Xi$ is a scalar auxiliary field,  
is invariant under the standard gauge transformations
for a 2-form,
\be
\d B= d \L,
\ee
and under the additional PST gauge transformations 
\be
\d B_{\m\n} =(\de_\m \Xi )\L_\n - (\de_\n \Xi )\L_\m
\ee
and 
\be
\d \Xi = \L \quad , \qquad \d B_{\m\n}=\frac{\L}{(\de \Xi )^2 } H^-_{\m\n\r}
\de^\r \Xi \quad .
\ee
Fixing the PST gauge in an appropriate way, 
one ends up with a self-dual 3-form, after
eliminating the auxiliary scalar. It should be 
observed that this gauge choice can not be
imposed directly on the Lagrangian, not defined if $\Xi=0$.

This method has already been applied to a number of systems, including 
$(1,0)$ six-dimensional supergravity 
coupled to tensor multiplets \cite{6bpst} and type-IIB 
ten-dimensional supergravity \cite{10bpst}, whose (local) gravitational anomaly 
has been shown to reproduce \cite{anomalypst} the 
well-known results \cite{agw} of Alvarez-Gaum\'e and Witten. Here we 
use the results of \cite{6bpst}, as well as those of \cite{rs1,fabio}, 
in which 
the same construction has been applied to the case with also vector 
multiplets and hypermultiplets.

We have a single self-dual 3-form and $n_T$ antiself-dual 
3-forms, where
$n_T$ is equal to 17 in the $T^4/Z_2$ model of \cite{bsb}. These forms
are obtained dressing with the scalars in the tensor multiplets the
3-forms
\be
H^r = d B^r - c^{rz} \w^{(9)z} \quad ,\label{cs6}
\ee
where the index $z$ runs over the various semi-simple factors of the 
gauge group in the 9-9 sector, $\w$ is the Chern-Simons 3-form and the $c$'s are constants
(we denote with $z=1$ the group under which the hypermultiplets are
charged). 
More precisely, the combinations
\be
H_{\m\n\r}=v_r H^r_{\m\n\r}
\ee
and
\be
H^M_{\m\n\r}= x^M_r  H^r_{\m\n\r}
\ee
are respectively self-dual and antiself-dual \cite{romans}, 
to lowest order in the fermi fields, although in the complete
lagrangian these conditions 
are modified by the inclusion of fermionic bilinears.
As in ten dimensions, the gauge invariance of $H^r$ in eq. (\ref{cs6}) implies
that $B^r$ vary as 
\be
\d B^r = c^{rz} tr_z ( \L^{(9)} d A^{(9)} ) 
\label{gaugeB}
\ee
under gauge transformations. 

To lowest order in the fermi fields, 
the Lagrangian describing the coupling of the supergravity  
multiplet to $n_T$ tensor multiplets, vector multiplets and $n_H$ hypermultiplets is
\bea
e^{-1}{\cal{L}}_{susy} = &-& \frac{1}{4}R +\frac{1}{12}G_{rs} H^{r \m\n\r} H^s_{\m\n\r}
+\frac{1}{4} g_{\bar{\a}\bar{\b}} \de_\m \Phi^{\bar{\a}}\de^\m \Phi^{\bar{\b}} 
-\frac{1}{2} v_r c^{rz} tr_z (F^{(9)}_{\m\n}F^{(9)\m\n}) \nonumber\\
&-& \frac{\de^\m \Xi \de^\s \Xi }{4(\de \Xi )^2 }
[{\cal H}^-_{\m\n\r} {\cal H}^-_{\s}{}^{\n\r} +
{\cal H}^{M+}_{\m\n\r} {\cal H}^{M+}{}_\s{}^{\n\r} ]
-\frac{1}{8e}\e^{\m\n\r\sigma\delta\tau}B^r_{\m\n}c_r^z tr_{z}
(F^{(9)}_{\r\sigma}
F^{(9)}_{\delta\tau}) \nonumber \\
&+& \frac{1}{2} g_{\a\b}(\phi ) D_\m \phi^\a D^\m \phi^\b 
+\frac{1}{4v_r c^{r1}} {\cal{A}}_\a^A{}_B 
{\cal{A}}_\b^B{}_A \xi^{\a i} \xi^{\b i} 
\nonumber \\
&-& \frac{i}{2}(\bar{\psi}_\m \g^{\m\n\r} D_\n 
\psi_\r )-\frac{i}{2}v_r H^{r \m\n\r}(\bar{\psi}_\m \g_\n \psi_\r)
+\frac{i}{2} (\bar{\chi}^M \g^\m D_\m \chi^M )\nonumber \\ 
&-&\frac{i}{24}v_r H^r_{\m\n\r} (\bar{\chi}^M \g^{\m\n\r}
\chi^M ) +\frac{1}{2}x^M_r \de_\n v^r (\bar{\psi}_\m \g^\n
\g^\m \chi^M) -\frac{1}{2} x^M_r H^{r \m\n\r} ( \bar{\psi}_\m
\g_{\n\r} \chi^M )\nonumber\\
&+& \frac{i}{2}(\bar{\Psi}_a \g^\m D_\m \Psi^a ) +\frac{i}{24}
v_r H^r_{\m\n\r} (\bar{\Psi}_a \g^{\m\n\r}\Psi^a ) - V_\a^{aA}D_\n \phi^\a
(\bar{\psi}_{\m A} \g^\n \g^\m \Psi_a )\nonumber \\
&+& i v_r c^{rz} tr_z(\bar{\l}^{(9)} \g^\m D_\m \l^{(9)} )
+\frac{i}{\sqrt{2}} v_r c^{rz} tr_z[F^{(9)}_{\n\r} (\bar{\psi}_\m \g^{\n\r}
\g^\m \l^{(9)} )]
\nonumber\\
&+& \frac{1}{\sqrt{2}} x^M_r c^{rz} tr_z [F^{(9)}_{\m\n} 
(\bar{\chi}^M \g^{\m\n}\l^{(9)} )]
-\frac{i}{12}c_r^z H^r_{\m\n\r} tr_z (\bar{\l}^{(9)} \g^{\m\n\r} \l^{(9)})
\nonumber\\
&-& \sqrt{2} V_\a^{aA} \xi^{\a i}(\bar{\l}^{(9)i}_A \Psi_a ) 
+\frac{i}{\sqrt{2}}{\cal{A}}_\a^A{}_B 
\xi^{\a i }(\bar{\l}^{(9)i}_A \g^\m \psi_\m^B ) \nonumber\\
&+& \frac{1}{\sqrt{2}}{\cal{A}}_\a^A{}_B 
\frac{x^M_r c^{r1}}{v_s c^{s1}}\xi^{\a i}
(\bar{\l}^{(9)i}_A \chi^{MB} ) \quad,
\label{pstlag}
\eea
where $G_{rs} = v_r v_s + x^M_r x^M_s$, while
\be
{\cal H}_{\m\n\r} = v_r {H}^{r}_{\m\n\r}
-\frac{3i}{2} (\bar{\psi}_{[\m} \g_\n \psi_{\r ]})
-\frac{i}{8} 
(\bar{\chi}^M \g_{\m\n\r} \chi^M )+\frac{i}{8} 
(\bar{\Psi}_a \g_{\m\n\r} \Psi^a)
\ee
and
\be
{\cal H}^M_{\m\n\r} = x^M_r {H}^{r}_{\m\n\r} 
+ \frac{3}{2} (\bar{\chi}^M\g_{[\m\n}
\psi_{\r ]}) +\frac{i}{4} x^M_r c^{rz} tr_z (\bar{\l}^{(9)} \g_{\m\n\r}
\l^{(9)} )
\ee
satisfy on-shell self-duality and antiself-duality conditions,
respectively.
Finally, $\Xi$ is the PST auxiliary field.

Due to eq. (\ref{gaugeB}), the Wess-Zumino term 
$B \wedge F \wedge F$ is not gauge invariant, and thus the variation 
of eq. (\ref{pstlag}) under gauge transformations produces the consistent
gauge anomaly
\be 
{\cal{A}}_\L =- \frac{1}{4} \e^{\m\n\r\sigma\delta\tau} c_r^z c^{rz^\prime} tr_z ({\L}^{(9)} 
\de_\m A^{(9)}_\n ) tr_{z^\prime} (F^{(9)}_{\r\sigma} F^{(9)}_{\delta\tau} )\quad ,
\label{consanomaly}
\ee 
related by the Wess-Zumino conditions to the supersymmetry anomaly
\be 
{\cal{A}}_\e = \e^{\m\n\r\sigma\delta\tau} c_r^z c^{r z^\prime}
[-\frac{1}{4} tr_z (
\delta_\e A^{(9)}_\m A^{(9)}_\n ) tr_{z^\prime} (F^{(9)}_{\r\sigma} 
F^{(9)}_{\delta\tau})  
-\frac{1}{6} tr_z (
\delta_\e A^{(9)}_\m F^{(9)}_{\n\r} ) \w^{(9)z^\prime}_{\sigma\delta\tau} ]\ ,
\label{susyanomaly}
\ee
that one can recover varying the Lagrangian of eq. (\ref{pstlag}) under
the supersymmetry transformations 
\bea
& & \delta e_\m{}^m = -i ( \bar{\e} \g^m \psi_\m ) \quad , \nonumber\\ 
& & \delta B^r_{\m\n} =i v^r ( \bar{\psi}_{[\m} \g_{\n]} \e )
+\frac{1}{2} x^{Mr} ( \bar{\chi}^M \g_{\m\n} \e ) 
- 2c^{rz}tr_z( A^{(9)}_{[\m} \delta A^{(9)}_{\n]} )
\quad , \nonumber\\ 
& & \d \Phi^{\bar{\a}} = V^{\bar{\a}M} (\bar{\chi}^M \e ) \quad , \nonumber\\
& & \delta \phi^\a = V^\a_{aA} ({\bar{\e}}^A \Psi^a ) \quad ,\nonumber\\
& & \d \Xi = 0 \quad , \nonumber \\
& & \delta A^{(9)}_\m = -\frac{i}{\sqrt{2}} (\bar{\e} \g_\m \l^{(9)} )\quad, 
\nonumber\\
& & \d \psi_\m^A =D_\m \e^A +
\frac{1}{4} K_{\m\n\r} \g^{\n\r} \e^A \quad , \nonumber \\
& & \d \chi^{MA} = -\frac{i}{2} V_{\bar{\a}}^M \de_\m \Phi^{\bar{\a}}\g^\m 
\e^A +\frac{i}{12}
K^M_{\m\n\r} \g^{\m\n\r} \e^A  \quad ,\nonumber \\
& & \delta \Psi^a = i \g^\m \e_A V_\a^{aA} D_\m \phi^\a \quad , \nonumber\\
& & \delta \l^{(9)A} = -\frac{1}{2\sqrt{2}} F^{(9)}_{\m\n} \g^{\m\n} \e^A \qquad 
\quad (z \neq 1 ) \quad , \nonumber \\
& & \delta \l^{(9)iA}= -\frac{1}{2\sqrt{2}} F^{(9)i}_{\m\n} \g^{\m\n} \e^A 
-\frac{1}{\sqrt{2}v_r c^{r1}} {\cal{A}}_\a^A{}_B \xi^{\a i} \e^B 
\label{susy6}
\eea
where
\be
K_{\m\n\r} ={\cal{H}}_{\m\n\r} -3 \frac{\de_{[\m} \Xi \de^\s  \Xi}{(\de \Xi )^2}
{\cal H}^-_{\s\n\r ]}
\ee
and
\be
K^M_{\m\n\r} ={\cal{H}}^M_{\m\n\r} -3 
\frac{\de_{[\m} \Xi \de^\s \Xi }{(\de \Xi)^2}
{\cal H}^{M+}_{\s\n\r ]}
\ee
are {\it identically} self-dual and antiself-dual, respectively.
In the complete theory, the anomalous terms would be exactly canceled by the 
anomalous contributions of fermion loops. 

Following the same reasoning as for 
the ten dimensional case of \cite{dm2}, we can describe the couplings to 
non-supersymmetric matter requiring that local supersymmetry be non-linearly 
realized on the ${\bar D}5$-branes, and requiring that the supersymmetry
variation of the non-supersymmetric fields be as in eq. (\ref{susygct}). 
As explained in \cite{dm2} and 
reviewed in the previous section, to lowest order in the fermions the
coupling between the supersymmetric sector and the non-supersymmetric
one is obtained dressing the bosonic fields in the supersymmetric sector with 
fermionic bilinears containing the goldstino, whose supersymmetry 
transformation is $\d \th =\e$ to lowest order in the fermionic fields. 
As a result, the supersymmetry variation of the dressed scalars in the tensor multiplets
\be
\hat{\Phi}^{\bar{\a}} = \Phi^{\bar{\a}} -V^{\bar{\a}M}(\bar{\th}
\chi^M ) +\frac{i}{24} V^{\bar{\a}M} x^M_r H^r_{\m\n\r}
(\bar{\th} \g^{\m\n\r} \th )
\ee
is a general coordinate transformation of parameter
\be
\xi_\m =-\frac{i}{2} (\bar{\th} \g_\m \e ) \quad .
\label{xi6}
\ee
This definition of $\hat{\Phi}$ then induces the corresponding dressing
\be
\hat{v}^r = v^r - x^{Mr} (\bar{\th} \chi^M ) -\frac{i}{24} H^r_{\m\n\r}
(\bar{\th} \g^{\m\n\r} \th ) \quad ,
\ee
and, in a similar fashion, the supersymmetry transformation of
\be
\hat{\phi}^\a = \phi^\a - V^\a_{aA} (\bar{\th}^A \Psi^a ) -\frac{i}{2}
V_{\b aA} V^{\a a B} (\bar{\th}^A \g^\m \th_B ) D_\m \phi^\b
\ee
is again a coordinate transformation with the same parameter, 
together with an additional gauge transformation of parameter
\be
\L^{(9)} = \xi^\m A^{(9)}_\m \quad .
\label{gauge9}
\ee 
Similarly, the supersymmetry variation of
\be
\hat{e}_\m{}^m  =  e_\m{}^m +i(\bar{\th}\g^m \psi_\m )-\frac{i}{2}
(\bar{\th} \g^m D_\m \th ) -\frac{i}{8} v_r H^r_{\m\n\r} 
(\bar{\th} \g^{m\n\r} \th ) 
\ee
contains also an additional
local Lorentz transformation of parameter
\be
\L^{mn} = -\xi^\m[ \w_\m{}^{mn} -v_r H^r_\m{}^{mn} ]
\label{lolo}
\ee
where $\w$ denotes the spin connection. 
Since the scalars in the non-supersymmetric 9-$\bar{5}$ sector are 
charged with respect to the vectors in the 9-9 sector, we define also
\bea
& & \hat{A}_\m^{(9)} = A^{(9)}_\m + \frac{i}{\sqrt{2}}(\bar{\th} \g_\m 
\l^{(9)} ) + \frac{i}{8} F^{(9)\n\r} (\bar{\th} \g_{\m\n\r} \th )\quad ( 
z \neq 1)\quad ,\nonumber \\
& & \hat{A}_\m^{(9)i} = A^{(9)i}_\m + \frac{i}{\sqrt{2}}(\bar{\th} \g_\m 
\l^{(9)i} ) + \frac{i}{8} F^{(9)i\n\r} (\bar{\th} \g_{\m\n\r} \th )
+\frac{i}{4 v_r c^{r1}}{\cal{A}}_\a^A{}_B \xi^{\a i} (\bar{\th}_A \g_\m 
\th^B ) \ ,
\eea
whose supersymmetry transformation is a general coordinate transformation of 
parameter as in eq. (\ref{xi6}), aside from a gauge transformation of parameter
as in (\ref{gauge9}).
If one requires that the supersymmetry variation of the vector $A^{(5)}_\m$ from the
non-supersymmetric $\bar{5}$-$\bar{5}$ sector be
\be
\d A^{(5)}_\m = F^{(5)}_{\m\n}\xi^\n \quad ,
\label{efxi}
\ee
namely a general coordinate transformation together with a gauge transformation of 
parameter
\be
\L^{(5)} = \xi^\m A^{(5)}_\m \quad ,\label{gauge5}
\ee
one obtains a supersymmetrization of the kinetic term for $A^{(5)}_\m$ writing
\be
 -\frac{1}{2} \hat{e} \hat{v}^r c_r^w tr_w (F^{(5)}_{\m\n} F^{(5)}_{\r\s} )
\hat{g}^{\m\r} \hat{g}^{\n\s} \quad ,\label{kin}
\ee
where
\be
\hat{g}_{\m\n} = \hat{e}_\m{}^m \hat{e}_{\n m} \quad ,
\ee
and the index $w$ runs over the various semi-simple factors of the gauge
group in the $\bar{5}$-$\bar{5}$ sector.  
In analogy with the ten-dimensional case, the uncanceled
NS-NS tadpole translates, in the 
low-energy theory, in the presence of a term 
\be
- \L \hat{e} f(\hat{\Phi}^{\bar{\a}}, {\hat\phi}^\a ) \quad ,\label{tadpole}
\ee
that depends on the scalars of the closed sector and contains the 
dilaton, that belongs to a hypermultiplet in type-I vacua. 
Thus, supersymmetry breaking 
naturally corresponds in this case also to a breaking of 
the isometries of the scalar manifolds.

Denoting with $S$ the scalars in the 9-$\bar{5}$ sector, charged with 
respect to the gauge fields in both the 9-9 and $\bar{5}$-$\bar{5}$ sectors, 
we define their covariant derivative as  
\be
\hat{D}_\m S = \de_\m S -i \hat{A}^{(9)}_\m  S  - i  {A}^{(5)}_\m  S 
\quad,
\ee  
so that the term
\be
\frac{1}{2} \hat{e}(\hat{D}_\m S )^\dagger (\hat{D}_\n S) \hat{g}^{\m\n}
\ee
is supersymmetric, if again the supersymmetry 
transformation of $S$ is a general
coordinate transformation, together with 
a gauge transformation of the right parameters.
As in the ten-dimensional case, if one considers terms up to quartic couplings
in the fermionic fields, one does not have to supersymmetrize terms that are 
quadratic in the additional fermions from the non-supersymmetric 
$\bar{5}$-$\bar{5}$ and $9$-$\bar{5}$ sectors. 
Denoting with $\l^{(5)}$ these fermions, the coupling of $\l^{(5)2}$
to the 3-forms is not determined by supersymmetry, and can only be determined 
by string considerations, as in \cite{dm2}. 

The inclusion of additional non-supersymmetric 
vectors modifies $H^r$, that now includes the Chern-Simons 3-forms 
corresponding to these fields, so that eq. (\ref{cs6}) becomes
\be
H^r = d B^r - c^{rz} \w^{(9)z} - c^{rw} \w^{(5)w} \quad .
\ee
The gauge invariance of $H^r$ then
requires that
\be
\d B^r = c^{rw} tr_w (\L^{(5)} d A^{(5)} ) \label{cs6bis}
\ee
under gauge transformations in the $\bar{5}$-$\bar{5}$ sector.
Consequently, the supersymmetry transformation of $B^r$ is 
also modified, and becomes
\bea
\delta B^r_{\m\n} &=& i v^r ( \bar{\psi}_{[\m} \g_{\n]} \e )
+\frac{1}{2} x^{Mr} ( \bar{\chi}^M \g_{\m\n} \e ) \nonumber\\ 
&-&2c^{rz}tr_z( A^{(9)}_{[\m} \delta A^{(9)}_{\n]} )
- 2c^{rw}tr_w( A^{(5)}_{[\m} \delta A^{(5)}_{\n]} )
\quad .
\eea
The complete reducible gauge anomaly
\bea 
{\cal{A}}_\L = & - & \frac{1}{4} \e^{\m\n\r\sigma\delta\tau}\{ c_r^z c^{rz^\prime} tr_z ({\L}^{(9)} 
\de_\m A^{(9)}_\n ) tr_{z^\prime} (F^{(9)}_{\r\sigma} F^{(9)}_{\delta\tau} )\nonumber\\
& + & c_r^z c^{rw} tr_z ({\L}^{(9)} 
\de_\m A^{(9)}_\n ) tr_{w} (F^{(5)}_{\r\sigma} F^{(5)}_{\delta\tau} )\nonumber\\
& + & c_r^w c^{rz} tr_w ({\L}^{(5)} 
\de_\m A^{(5)}_\n ) tr_{z} (F^{(9)}_{\r\sigma} F^{(9)}_{\delta\tau} )\nonumber\\
& + & c_r^w c^{rw^\prime} tr_w ({\L}^{(5)} 
\de_\m A^{(5)}_\n ) tr_{w^\prime} (F^{(5)}_{\r\sigma} F^{(5)}_{\delta\tau} ) \} \quad ,
\label{complconsanomaly}
\eea 
related by the Wess-Zumino conditions to the supersymmetry anomaly
\bea 
{\cal{A}}_\e & = & \e^{\m\n\r\sigma\delta\tau} \{ c_r^z c^{r z^\prime}
[-\frac{1}{4} tr_z (
\delta_\e A^{(9)}_\m A^{(9)}_\n ) tr_{z^\prime} (F^{(9)}_{\r\sigma} 
F^{(9)}_{\delta\tau})  
-\frac{1}{6} tr_z (
\delta_\e A^{(9)}_\m F^{(9)}_{\n\r} ) \w^{(9)z^\prime}_{\sigma\delta\tau} ]\nonumber \\
& + & c_r^z c^{r w}
[-\frac{1}{4} tr_z (
\delta_\e A^{(9)}_\m A^{(9)}_\n ) tr_{w} (F^{(5)}_{\r\sigma} 
F^{(5)}_{\delta\tau})  
-\frac{1}{6} tr_z (
\delta_\e A^{(9)}_\m F^{(9)}_{\n\r} ) \w^{(5)w}_{\sigma\delta\tau} ]\nonumber \\
& +& c_r^w c^{r z}
[-\frac{1}{4} tr_w (
\delta_\e A^{(5)}_\m A^{(5)}_\n ) tr_{z} (F^{(9)}_{\r\sigma} 
F^{(9)}_{\delta\tau})  
-\frac{1}{6} tr_w (
\delta_\e A^{(5)}_\m F^{(5)}_{\n\r} ) \w^{(9)z}_{\sigma\delta\tau} ]\nonumber \\
& +& c_r^w c^{r w^\prime}
[-\frac{1}{4} tr_w (
\delta_\e A^{(5)}_\m A^{(5)}_\n ) tr_{w^\prime} (F^{(5)}_{\r\sigma} 
F^{(5)}_{\delta\tau})  
-\frac{1}{6} tr_w (
\delta_\e A^{(5)}_\m F^{(5)}_{\n\r} ) \w^{(5)w^\prime}_{\sigma\delta\tau} ] \} \ ,
\label{complsusyanomaly}
\eea
is induced by the Wess-Zumino term
\be
-\frac{1}{8} \e^{\m\n\r\s\d\t}B^r_{\m\n} c_{r}^w tr_w (F_{\r\s}^{(5)} F_{\d\t}^{(5)})\quad .
\label{gs6}
\ee
It should be noticed that, as in the case of linearly realized supersymmetry, 
eqs. (\ref{complconsanomaly}) and (\ref{complsusyanomaly}) satisfy
the Wess-Zumino condition
\be
\d_\L {\cal{A}}_\e = \d_\e {\cal{A}}_\L \quad ,
\ee
since the explicit form of the gauge field supersymmetry variation 
plays no role in its proof. We expect that, to higher order in the fermions, 
the supersymmetry anomaly will be modified by gauge-invariant terms 
as in \cite{ggs3,fabio}.
From the definition of $H^r$, one can deduce the Bianchi identities  
\be
\de_{[\m} H^r_{\n\r\s ]} =-\frac{3}{2} c^r_z tr_z (F^{(9)}_{[\m\n} F^{(9)}_{\r\s ]} ) 
-\frac{3}{2} c^r_w tr_w (F^{(5)}_{[\m\n} F^{(5)}_{\r\s ]} )\quad .
\ee

We now want to determine the terms proportional
to $F \wedge F$ containing the goldstino 
that one has to add for the consistency of the model.
Unlike the ten dimensional case, where duality maps the 2-form theory 
with Chern-Simons couplings
to the 6-form theory with Wess-Zumino couplings, in this case 
the low-energy effective action contains both Chern-Simons and Wess-Zumino  
couplings.
First of all, we observe that for the quantity
\bea
\hat{B}^r_{\m\n} &=& B^r_{\m\n} -iv^r (\bar{\psi}_{[\m}\g_{\n ]}\th )
-\frac{1}{2} x^{Mr} (\bar{\chi}^M \g_{\m\n} \th ) -\frac{2i}{\sqrt{2}}c^{rz}tr_z 
[A_{[\m}^{(9)} (\bar{\th} \g_{\n ]}\l^{(9)} ) ]\nonumber \\
&+& \frac{i}{8} (\de_\r v^r ) (\bar{\th}\g_{\m\n}{}^\r \th )
+\frac{i}{8} x^{Mr} H^M_{[\m}{}^{\r\s} (\bar{\th} \g_{\n ] \r\s} \th )+
\frac{i}{2} v^r (\bar{\th} \g_{[\m }D_{\n ]}\th )\nonumber \\
&-& \frac{i}{4}c^{rz}tr_z [A_{[\m}^{(9)} F^{(9)\r\s} ](\bar{\th} \g_{\n ] \r\s} \th )
-\frac{i c^{r1}}{4 v_s c^{s1}} {\cal A}_\a^A{}_B \xi^{\a i} A_{[ \m}^{(9)i}
(\bar{\th}^A \g_{\n ]} \th^B ) 
\label{bextra}
\eea
the supersymmetry variation is a general coordinate 
transformation of the correct  
parameter, together with an additional tensor gauge transformation 
of parameter
\be
\L^r_\m = -\frac{1}{2} v^r \xi_\m -\xi^\n B^r_{\m\n} \quad ,
\label{tega}
\ee
as well as PST gauge transformations of parameters 
\be
\L^{(PST)r}_\m =\frac{\de^\s \Xi}{(\de \Xi )^2 }[v^r v_s H^{s-}_{\s\m\r}
- x^{Mr} x^M_s H^{s+}_{\s\m\r}] \xi^\r 
\label{pst1}
\ee
and 
\be
\L^{(PST)} = \xi^\m \de_\m \Xi 
\label{pst2}
\ee
and gauge transformations of the form  (\ref{gaugeB}) and 
(\ref{cs6bis}) whose parameters are
as in eq. (\ref{gauge9}) and (\ref{gauge5}). 
We should now consider all the terms proportional to $F \wedge F$
that arise, those directly introduced by 
the inclusion of the Chern-Simons 3-form for the fields in the $\bar{5}$-$\bar{5}$ sector, those originating from 
the consequent modification of the Bianchi identities,
and finally those introduced by the variation of the Wess-Zumino term.

The end result is that the variation of all these contributions gives
\bea
\d {\cal L} &=& \e^{\m\n\r\s\d\t} \{ -\frac{i}{4} v_r (\bar{\e} \g_\m \psi_\n )
 + \frac{1}{8} x^{M}_{r}(\bar{\e}\g_{\m\n} \chi^M ) \} c^{rw}tr_w (F^{(5)}_{\r\s} F^{(5)}_{\d\t})
\nonumber\\
&-& 2v_r c^{rw} tr_w (\d A^{(5)}_\m F^{(5)}_{\n\r} )K^{\m\n\r}-2 x^M_r c^{rw} tr_w
(\d A^{(5)}_\m F^{(5)}_{\n\r} )K^{M \m\n\r}  \quad .\label{deltalag}
\eea
The first two terms are canceled by the goldstino variation in the
additional couplings
\be
{\cal L}^\prime = \e^{\m\n\r\s\d\t} 
\{ \frac{i}{4}v_r(\bar{\th} \g_\m \psi_\n )
- \frac{1}{8} x^{M}_{r}(\bar{\th}\g_{\m\n} \chi^M ) \} c^{rw}tr_w (F^{(5)}_{\r\s} F^{(5)}_{\d\t})
\quad ,
\ee
where, however, the variations of the gravitino and the tensorinos 
produce additional terms. Some of these 
cancel the last two terms in eq. (\ref{deltalag}), while the remaining ones
are canceled by the goldstino variation in
\bea
{\cal L}^{\prime \prime} &=& \e^{\m\n\r\s\d\t} \{ -\frac{i}{32}\de^\r v_r
(\bar{\th}\g_{\m\n\r} \th )  -\frac{i}{8} v_r (\bar{\th} \g_\m
D_\n \th )  \nonumber \\
&-& \frac{i}{32}x^M_r K^M_\m{}^{\a\b} (\bar{\th}\g_{\n\a\b} \th ) 
\} c^{rw} tr_w (F^{(5)}_{\r\s} F^{(5)}_{\d\t} ) \quad .
\eea
If one restricts the attention to terms up to quartic fermion couplings, 
no further contributions are produced.
We can thus conclude that the non-linear realization of supersymmetry is 
granted by the inclusion of ${\cal L}^\prime$ and ${\cal L}^{\prime\prime}$ 
in the low-energy effective action. From eq. (\ref{bextra}) we also
see that these two contributions can be written in the compact form
\be
{\cal L}^\prime + {\cal L}^{\prime\prime}= -\frac{1}{4}\e^{\m\n\r\s\d\t}
B^r_{\m\n}{}^{extra} c_r^{w} tr_w (F^{(5)}_{\r\s} F^{(5)}_{\d\t} )\quad ,\label{2gs}
\ee
where
\bea
B^r_{\m\n}{}^{extra} = &-& iv^r (\bar{\psi}_{[\m}\g_{\n ]}\th )
-\frac{1}{2} x^{Mr} (\bar{\chi}^M \g_{\m\n} \th ) -\frac{2i}{\sqrt{2}}c^{rz}tr_z 
[A_{[\m}^{(9)} (\bar{\th} \g_{\n ]}\l^{(9)} ) ]\nonumber \\
&+&  \frac{i}{8} (\de_\r v^r ) (\bar{\th}\g_{\m\n}{}^\r \th )
+\frac{i}{8} x^{Mr} K^M_{[\m}{}^{\r\s} (\bar{\th} \g_{\n ] \r\s} \th )+
\frac{i}{2} v^r (\bar{\th} \g_{[\m }D_{\n ]}\th )
\eea
coincides with the counterterm of $B^r$ only if no 9-9 vectors are present.

We now want to interpret these non-geometric terms along the lines
of Section 3. To this end, observe that, if no 9-9 vectors are 
present, eq. (\ref{2gs}) is exactly twice the 
term that one should add to eq. (\ref{gs6}) in order to geometrize the 
Wess-Zumino term, substituting $B$ with $\hat{B}$. 
This means, roughly speaking, that half of the contribution in eq. (\ref{2gs})
comes from the Green-Schwarz term, and half from the Chern-Simons couplings. 
This interpretation is in perfect agreement with self-duality, and thus in six 
dimensions there is no duality transformation that can give a fully geometric 
Lagrangian. If also 9-9 vectors are in the spectrum, no additional terms are
produced in the lagrangian, in agreement with the fact that the additional terms of $\hat{B}^r$ in eq. 
(\ref{bextra}) are not gauge invariant.

To resume, the Lagrangian for supergravity coupled to tensor multiplets, 
hypermultiplets and non-supersymmetric vectors is
\bea
{\cal L} &=& {\cal L}_{susy} -\frac{1}{2} \hat{e}\hat{v}^r c_r^w tr_w (F^{(5)}_{\m\n} F^{(5)}_{\r\s} )
\hat{g}^{\m\r} \hat{g}^{\n\s} - \L \hat{e}f(\hat{\Phi}^{\bar{\a}}, \hat{\phi}^\a )\nonumber \\
&+&  \frac{1}{2} \hat{e}(\hat{D}_\m S )^\dagger ( \hat{D}_\n S ) \hat{g}^{\m\n}
-\frac{1}{8} \e^{\m\n\r\s\d\t} B^r_{\m\n} c_r^w tr_w (F^{(5)}_{\r\s}F^{(5)}_{\d\t}) \nonumber \\
&-& \frac{1}{4}\e^{\m\n\r\s\d\t}
B^r_{\m\n}{}^{extra} c_r^w tr_w (F^{(5)}_{\r\s} F^{(5)}_{\d\t} ) \quad . 
\label{complag6}
\eea
Since the supersymmetry transformation of other non-supersymmetric
fermions is of higher order in the fermi fields, at this level we can always
add them in the construction, while 
the couplings that can not be determined by supersymmetry could in 
principle be determined by string inputs, as in \cite{dm2}.

Finally, it is important to observe that without 9-9 vectors, 
although the Lagrangian 
(\ref{complag6}) is not completely geometric, the corresponding 
field equations are. 
Indeed, if one fixes the PST gauge in such a way 
that the 3-forms satisfy the standard (anti)self-duality conditions, 
the equation for the vector fields, up to terms quartic in the fermions, 
is 
\bea
& & \hat{e} D_\n [ \hat{v}^r c_{rw} F^{(5)}_{\r\s} \hat{g}^{\m\r}
\hat{g}^{\n\s}
]+\frac{1}{6} \e^{\m\n\r\s\d\t} 
c_{rw} F^{(5)}_{\n\r}
\hat{H}^r_{\s\d\t} \nonumber\\
& & + \frac{1}{12}\e^{\m\n\r\s\d\t} c_{rw} F^{(5)}_{\n\r} c^{rw^\prime}
\w^{(5)w^\prime}_{\s\d\t} 
+ \frac{1}{8}\e^{\m\n\r\s\d\t} c_{rw} A^{(5)}_{\n} c^{rw^\prime}
tr_{w^\prime}(F^{(5)}_{\r\s} F^{(5)}_{\d\t} ) =0 \quad ,
\eea
where 
\be
\hat{H}^r_{\m\n\r} = 3 \de_{[\m} \hat{B}^r_{\n\r ]} - c^r_w \w^{(5)w}_{\m\n\r} 
\quad ,
\ee
and this is nicely of geometric form.
It should be noticed that no additional counterterms 
containing the goldstino have to be added if also 9-9 vectors are present.
In fact, all the terms in $\hat{B}^r$ induced by $A^{(9)}$ are 
not gauge invariant,
and their inclusion in the lagrangian is forbidden because it would modify
the gauge anomaly.  The resulting equation for $A^{(5)}$ is then
\bea
& & \hat{e} D_\n [ \hat{v}^r c_{rw} F^{(5)}_{\r\s} \hat{g}^{\m\r}
\hat{g}^{\n\s}
]+\frac{1}{6} \e^{\m\n\r\s\d\t} 
c_{rw} F^{(5)}_{\n\r}
\tilde{H}^r_{\s\d\t} \nonumber\\
& & + \frac{1}{12}\e^{\m\n\r\s\d\t} c_{rw} F^{(5)}_{\n\r} c^{rz}
\w^{(9)z}_{\s\d\t} 
+ \frac{1}{8}\e^{\m\n\r\s\d\t} c_{rw} A^{(5)}_{\n} c^{rz}
tr_{z}(F^{(9)}_{\r\s} F^{(9)}_{\d\t}) \nonumber \\
& & + \frac{1}{12}\e^{\m\n\r\s\d\t} c_{rw} F^{(5)}_{\n\r} c^{rw^\prime}
\w^{(5)w^\prime}_{\s\d\t} 
+ \frac{1}{8}\e^{\m\n\r\s\d\t} c_{rw} A^{(5)}_{\n} c^{rw^\prime}
tr_{w^\prime}(F^{(5)}_{\r\s} F^{(5)}_{\d\t} ) =0 \quad ,
\label{eqcons}
\eea
where
\be
\tilde{H}^r_{\m\n\r} = 3 \de_{[\m} B^r_{\n\r ]} + 3 \de_{[\m} B^r_{\n\r ]}{}^{extra}
- c^{rz} \w^{(9)z}_{\m\n\r}- c^{rw} \w^{(5)w}_{\m\n\r}
\ee
is geometric up to gauge-invariant terms proportional to $c^{rz}$.
The result is thus in agreement with what expected
by anomaly considerations.  If gauge and supersymmetry anomalies are 
absent, the $A^{(5)}$ equation is mapped into itself by supersymmetry:
this is the very reason why this equation is geometric.  In the presence of
gauge and supersymmetry anomalies, as long as 9-9 vectors are absent, 
the equation
for $A^{(5)}$ is still geometric, albeit not gauge invariant.  
The supersymmetry anomaly, in this case, results from the gauge 
transformation contained in eq. (\ref{efxi}).
When also 9-9 vectors are present,  these arguments do not apply, 
and thus in eq. (\ref{eqcons}) 
the geometric structure is violated 
by terms proportional to $c^{rz} c_{r}^{w}$.  

The consistent formulation described above 
can be reverted to a supersymmetric formulation in terms of 
covariant non-integrable field equations \cite{ggs,rs2}, 
that embody the corresponding covariant gauge anomaly
\bea
{\cal A}_\L^{cov} & = & \frac{1}{2} \e^{\m\n\r\s\d\t} [c^{rz} c_r^{z^\prime} tr_z (\L^{(9)} F^{(9)}_{\m\n} )
tr_{z^\prime} (F^{(9)}_{\r\s} F^{(9)}_{\d\t} )\nonumber \\
& + & c^{rz} c_r^{w} tr_z (\L^{(9)} F^{(9)}_{\m\n} )
tr_{w} (F^{(5)}_{\r\s} F^{(5)}_{\d\t} )\nonumber \\
& + & c^{rw} c_r^{z} tr_w (\L^{(5)} F^{(5)}_{\m\n} )
tr_{z} (F^{(9)}_{\r\s} F^{(9)}_{\d\t} )\nonumber \\
& + & c^{rw} c_r^{w^\prime} tr_w (\L^{(5)} F^{(5)}_{\m\n} )
tr_{w^\prime} (F^{(5)}_{\r\s} F^{(5)}_{\d\t} ) ]\quad ,
\label{covan}
\eea
given by the divergence of the covariant equation for $A_\m^{(5)}$,
\be
\hat{e} D_\n [ \hat{v}^r c_{rw} F^{(5)}_{\r\s}
\hat{g}^{\m\r} \hat{g}^{\n\s} ]+\frac{1}{6} \e^{\m\n\r\a\b\g} c_{rw} F^{(5)}_{\n\r}
\tilde{H}^r_{\a\b\g} =0 \quad ,
\label{coveq}
\ee
and the divergence of the covariant equation for $A_\m^{(9)}$ \cite{rs2}.
Without 9-9 vectors, eq. (\ref{coveq}) is both geometric and 
gauge-covariant, while, if 9-9 vectors are present, 
the geometric structure is violated 
by gauge-covariant terms proportional again to $c^{rz} c_{r}^{w}$.

\section{Conclusions}
In this paper we have extended the results of \cite{dm2} on the
low-energy effective action for models with brane supersymmetry
breaking. In this class of models a supersymmetric bulk is
coupled to a non-supersymmetric open sector, and as a result
local supersymmetry is non-linearly realized {\it \`a la}
Volkov-Akulov. In particular, we have shown that, up to quartic order
in the fermions, the low-energy
couplings between the supersymmetric bulk and the non-supersymmetric open
sector in the ten-dimensional $USp(32)$ model of \cite{sugimoto} are {\it all}
of geometric origin, being induced by the dressing of the bulk fields
in terms of the goldstino, provided one turns to the dual 6-form
\cite{chamseddine} formulation. Thus, in retrospect, the non-geometric
terms in \cite{dm2} are precisely what is needed to geometrize the dual
form of the theory, where the (high-derivative)
Chern-Simons couplings are absent. We have completed a similar construction
for six-dimensional models with brane supersymmetry breaking. Since in
this case both Chern-Simons and Wess-Zumino terms are simultaneously
present, not all couplings in the Lagrangian can be related to 
goldstino-dependent dressings of bulk fields. 
However, in the absence of supersymmetric vectors, the 
field equations exhibit this geometric structure, that is
naturally violated in the general case by anomalous terms.  

It would be interesting to apply the same construction to the four-dimensional 
brane supersymmetry breaking models of \cite{bsb}, and in general to brane-world scenarios (see 
\cite{dudrev} and references therein) in which 
supersymmetry is linearly realized in the gravitational sector and non-linearly
realized in the brane universe.  However,  
in four dimensions the gravitino can acquire a Majorana
mass through a super-Higgs mechanism \cite{suhig}, 
and it is this difference with respect to minimal 
supergravity in ten and six dimensions that makes the models studied 
in this paper rather peculiar.

A general property of all this class of models is the presence of 
a dilaton tadpole of positive sign, required in order 
to have a correct kinetic term for the goldstino \cite{dm2}, and
guaranteed by the residual tension of anti-branes and orientifolds. 
In ten dimensions the tadpole signals the impossibility of having
maximally symmetric vacuum configurations \cite{dm3}, and one should try 
to analyze the same effects in the six-dimensional models discussed in 
this paper.

Finally, it should be observed that it is always possible, in any 
supersymmetric theory coupled to a goldstino, to dress the fields in 
the linear sector by terms containing the goldstino itself.
This is a property of the commutator of two supersymmetries:
by construction, a supersymmetry transformation on the dressed fields 
exactly corresponds to the commutator of two supersymmetries on
the linear fields, producing general coordinate transformations together 
with all the other local symmetry transformations. 
In the six dimensional case discussed in Section $4$, this can be 
explicitly verified:
the parameters of eqs. (\ref{xi6}), (\ref{gauge9}), (\ref{lolo}), 
(\ref{tega}), (\ref{pst1}) and (\ref{pst2}) coincide with those
coming from the supersymmetry algebra, provided one
substitutes  $ -\frac{i}{2}(\bar{\th} \g_\m \e )$ with 
$- i (\bar{\e}_1 \g_\m \e_2 )$ \cite{ggs3,fabio,rs1}.
Following this way of reasoning, one could try to generalize the results 
obtained here to all orders in the fermi fields.

\section*{Acknowledgements}

It is a pleasure to thank E. Dudas, M. Larosa, J. Mourad, Ya.S. Stanev 
and in particular 
A. Sagnotti, for many illuminating discussions.  This work was supported 
in part by the EEC contracts HPRN-CT-2000-00122 and HPRN-CT-2000-00148, 
and by the INTAS contract 99-1-590.

\end{document}